\documentclass[12pt, draftclsnofoot,onecolumn]{IEEEtran}

\usepackage{dsfont}
\usepackage{amssymb}
\usepackage{float}
\usepackage{subfig}
\usepackage{graphicx, amssymb, amsmath}
\usepackage{extarrows}
\usepackage{algorithm} 
\usepackage{algorithmic} 
\usepackage{multirow} 
\usepackage{color}
\usepackage{cite}
\emergencystretch=\maxdimen
\IEEEoverridecommandlockouts
\begin{document}

\title{AoI-Delay Tradeoff in Mobile Edge Caching with Freshness-Aware Content Refreshing}

\author{Shan~Zhang,~\IEEEmembership{Member,~IEEE,}
	Liudi~Wang,~\IEEEmembership{Student~Member,~IEEE,}
	Hongbin~Luo,~\IEEEmembership{Member,~IEEE,}
	Xiao~Ma,~~\IEEEmembership{Member,~IEEE,}
	and~Sheng~Zhou,~\IEEEmembership{Member,~IEEE,}
	\thanks{Shan~Zhang, Liudi~Wang, and Hongbin~Luo are with Beijing Key Laboratory of Computer Networks, the School of Computer Science and Engineering, Beihang University, Beijing, China, 100191 (email: \{zhangshan18, wangliudi, luohb\}@buaa.edu.cn).}
	\thanks{Xiao Ma is with Beijing University of Posts and Telecommunications, Beijing, China, 100876 (email: maxiao18@bupt.edu.cn).}
	\thanks{Sheng Zhou is with the Department of Electronic Engineering, Tsinghua University, Beijing, China, 100084 (email: sheng.zhou@tsinghua.edu.cn).}
	\thanks{This work has been submitted to the IEEE for possible publication.  Copyright may be transferred without notice, after which this version may no longer be accessible. Part of this work has been presented in the IEEE GLOBECOM 2019 \cite{mine_GC19_AOI}.}.
}

\maketitle

\begin{abstract}
	
	Mobile edge caching can effectively reduce service delay but may introduce information staleness, calling for timely content refreshing.
	However, content refreshing consumes additional transmission resources and may degrade the delay performance of mobile systems.
	In this work, we propose a freshness-aware refreshing scheme to balance the service delay and content freshness measured by Age of Information (AoI).
	Specifically, the cached content items will be refreshed to the up-to-date version upon user requests if the AoI exceeds a certain threshold (named as refreshing window).
	The average AoI and service delay are derived in closed forms approximately, which reveals an AoI-delay tradeoff relationship with respect to the refreshing window.
	In addition, the refreshing window is optimized to minimize the average delay while meeting the AoI requirements, and the results indicate to set a smaller refreshing window for the popular content items.
	Extensive simulations are conducted on the OMNeT++ platform to validate the analytical results.
	The results indicate that the proposed scheme can restrain frequent refreshing as the request arrival rate increases, whereby the average delay can be reduced by around 80\% while maintaining the AoI below one second in heavily-loaded scenarios.
	
\end{abstract}

\begin{IEEEkeywords}
	mobile edge caching, age of information (AoI), delay, cache refreshing, content dynamics
\end{IEEEkeywords}

\section{Introduction}
	
	Mobile edge caching enables content service in proximity by using the storage resources of radio access facilities and mobile devices, bringing tremendous benefits for both network users and operators \cite{Bastug14_cache_framework_BS_D2D_mag}.
	The in-proximity service can effectively reduce the end-to-end delay and better support time-critical applications \cite{Vu18_cache_performance_TWC,Xiao_TCC,cheng2019space}.
	Besides, the backhaul transmission pressure is relieved with reduced duplicated transmissions, whereby the capacity can be effectively enhanced especially in the backhaul-constrained dense networks \cite{Wang18_cache_backhaul_TWC}.
	Furthermore, assisted with big data analysis, the contents can be cached pro-actively based on user preference, mobility and behaviors, providing fine-grained customized services \cite{mine_JSAC_VNET,Shen20_Survey_AI_slicing}.	
	With these attractive potential benefits, mobile edge caching is considered as a cornerstone of the 5G and beyond networks \cite{GaoZhen_next_generation,Zhuang20_Proceeding}.
	
	Despite the attractive benefits, mobile edge caching may lead to staleness of dynamic items whose content information changes with time and environment \cite{Wang14_cache_framework_wireless_mag}.
	Examples can be the content of the same URL, real-time street maps, traffic congestion of a certain region, temperature and air conditions of a room, etc. 
	Therefore, the cached items should be refreshed timely to the most recent versions.
	However, cache refreshing introduces additional transmissions, which can degrade the delay performance of content delivery due to the constrained bandwidth resources \cite{mine_GC19_AOI}.
	Therefore, efficient cache refresh schemes should be devised to optimize both delay and content freshness.
	A body of works have been conducted on mobile edge caching deployment and management, but most of them focus on static content items which require no refreshing \cite{YangPeng20_TII}.
	Early studies have proposed cache refreshing schemes for the database in wired networks, whereas these schemes are not applicable for mobile edge caching due to the limited bandwidth and unreliable wireless transmissions \cite{Si97_adaptive_refresh}.

	In this work, we propose a freshness-aware content refreshing scheme for cache-enabled mobile networks, in order to minimize the average service delay while guaranteeing content freshness.
	Age of Information (AoI) is adopted to characterize content freshness, which is the time elapsed since the generation of the current version \cite{Kaul11_AOI_concept_conf}.
	A BS caches content items collected from source nodes to serve mobile users on demand, both through wireless transmissions.
	The BS always checks if the AoI of a cached item exceeds a certain threshold (defined as refreshing window) or not before delivering it to mobile users.
	The BS will directly deliver the cached version if still fresh.	
	Otherwise, the BS will first fetch the latest version for refreshing, and then deliver it to mobile users.
	
	As the system bandwidth is constrained, there exists a tradeoff between AoI and delay with respect to the refreshing window size, which is investigated in an analytical way.
	However, the analysis of AoI and delay is challenging due to the coupling effect of multi-content refresh and request arrival, multi-dimensional randomness of traffic arrival and wireless transmissions.
	To address these issues, we first study the BS service process in the single-source scenario, whereby the average AoI and service delay are derived in closed forms approximately based on queuing models and analysis.
	The results show that the average AoI increases with the refreshing window size in a convex manner and demonstrates an asymptotically linear relationship as the refreshing window size increases.
	On the contrary, the average delay is proved to decrease with the refreshing window convexly, revealing a tradeoff with the AoI.	
	Furthermore, the proposed scheme restrains the BS from frequently refreshing an item when the request arrival rate increases, saving more bandwidth for content delivery to mobile users.
	Then, the results are further extended to the multi-source scenario, whereby the refreshing window is optimized for each individual content item to minimize the average delay while meeting the average AoI requirement.
	The problem is proved to be convex and numerical results can be obtained through MATLAB toolboxes.	
	Extensive simulations are conducted on the OMNeT++ platform to validate the theoretical analysis.
	In addition, the performance of the proposed scheme is also compared with the conventional eager refreshing scheme where the items are always refreshed before delivery.
	The results show that the proposed scheme can effectively reduce the service delay and enhance the service capability by avoiding frequent cache refreshing, especially in heavy-loaded scenarios.
	In particular, the average delay can be reduced by around 80\% while maintaining the AoI below one second, according to the real-trace experiments.		
	
	The main contributions of this work are as follows:
	\begin{itemize}
		\item A refreshing scheme is devised for mobile edge caching considering the dynamic variation of content information, to guarantee the freshness of user-received contents;
		\item The average AoI and delay performances are analyzed theoretically, revealing a tradeoff with respect to the refreshing window;
		\item The proposed scheme is optimized for the multi-source scenario, where the optimal refreshing window provides insights into content freshness management of practical mobile edge caching systems.
	\end{itemize}
	
	The remaining of this paper is organized as follows. 
	The existing works on cache management are reviewed in Section~\ref{sec_review}, whereby the novelty of this work is highlighted.
	Section~\ref{sec_system_model} builds the system model, and the freshness-aware content refreshing scheme is proposed.
	The average AoI and delay are analyzed in Section~\ref{sec_single_source}, for the single-source scenario.
	Then, Section~\ref{sec_multi_source} extends the results to the multi-source scenario, based on which the refreshing window is optimized.
	Simulation results are provided in Section~\ref{sec_simulation}, followed by the conclusions in Section~\ref{sec_conclusions}.

\section{Literature Review}
	\label{sec_review}
	Focusing on the conflict of explosive contents and constrained storage resources, extensive efforts have been denoted to enhancing the caching efficiency from the aspects of where to cache (i.e., cache deployment) \cite{mine_hierarchical_cache_TVT, Kwak18_hybrid_cache_TWC}, what to cache (i.e., content placement) \cite{YangPeng19_TMM,Liu16_EE_cache_JSAC,mine_cache_TMC,Xiao_Infocom20}, and how to cache (i.e., cache update) \cite{Gao19_probabilistic_cache_1, Gao19_probabilistic_cache_2, Garetto15_analysis_cache_update_stationary_infocom, Hsu16_cache_time_spread_ICC,Ahlehagh14_adaptive_caching_TON, Gao20_probabilistic_cache_TMC, Müller17_learn_cache_bandit_TWC}.
	Cache deployment is usually conducted in long-time scale with network planning, which determines where to deploy the cache instances under constrained budget.
	In this regard, the deployment costs of different network entities (like the remote servers, gateways, and heterogeneous BSs) should be considered \cite{mine_hierarchical_cache_TVT,Kwak18_hybrid_cache_TWC}.
	Content placement selects the items to cache based on the content popularity and user requests, aiming at different design objectives such as hit rate maximization, delay minimization, service experience enhancement, mobility support \cite{Liu16_EE_cache_JSAC,mine_cache_TMC,Xiao_Infocom20}.
	Cache update deals with the popularity variation in spatial or temporal domains.
	Specifically, the items with faded popularity are discarded to make room for new popular ones, so as to maintain the content hit rate \cite{Gao19_probabilistic_cache_1,Gao19_probabilistic_cache_2}.
	With perfect knowledge of content popularity, existing works have proposed to update the cache during off-peak hours, by utilizing the idle transmission resources opportunistically \cite{Garetto15_analysis_cache_update_stationary_infocom, Hsu16_cache_time_spread_ICC}.
	Adaptive content update schemes remove the least used items out of cache, which can adapt to the popularity variations without perfect knowledge of content popularity \cite{Ahlehagh14_adaptive_caching_TON, Gao20_probabilistic_cache_TMC}.
	Furthermore, online update schemes have been also proposed based on machine learning methods such as the contextual multi-armed bandit optimization, where the BSs update the cached items based on the regular observation of content hit rate \cite{Müller17_learn_cache_bandit_TWC}.
	Although insightful, the-state-of-the-art studies on mobile edge caching mostly focus on static content items, wherein the content of an item is assumed to be unchanged.
	
	Considering the dynamics of contents, the freshness or timeliness has appeared as an importance performance metric, raising the new concept of AoI \cite{Kaul12_AoI_basic_update_infocom}. 
	The theoretical results have shown that the conventional delay-optimal transmission strategies may not be AoI-optimal \cite{Najm16_AoI_performance_gamma_ISIT}.
	Accordingly, effective transmission strategies are devised to reduce the peak or average AoI at the receiver side in different scenarios  \cite{SunJingzhou_TCOM_AoI,Zhou19_AoI_IoT_TWC,Jiang_AoI,Chen20_AOI_VANET_TWC}.
	The very recent works have introduced content freshness metrics into mobile edge caching \cite{Kam17_AoI_cache_fresh_popular_ISIT,Yates17_AoI_cache_update_17,Zhong18_cache_update_AoI_ISIT}.
	Considering that the content popularity can fade with time, Kam \emph{et. al} have proposed to predict content popularity based AoI and request rate \cite{Kam17_AoI_cache_fresh_popular_ISIT}.
	However, the proposed cache update scheme still aims at maximizing the hit rate instead of maintaining content freshness \cite{Kam17_AoI_cache_fresh_popular_ISIT}.
	Cache refreshing schemes have been devised to minimize the average AoI of a local cache system, considering the constrained transmission resources in \cite{Yates17_AoI_cache_update_17} and \cite{Zhong18_cache_update_AoI_ISIT}.
	Notice that these works focus on the AoI of cached items, whereas the freshness of user-received contents can be more important to the quality of experience.
	
	To summarize, the novelty of this work is two-fold compared with existing works: (1) a freshness-aware content refreshing scheme is proposed and optimized to guarantee the freshness of user received contents; and (2) the interplay between the average service delay and the average AoI of user-received contents is revealed through both theoretical analysis and simulations.

\section{System Model for Content Refreshing}
    \label{sec_system_model}

    \begin{figure}[!t]
    	\centering
    	\includegraphics[width=3.5in]{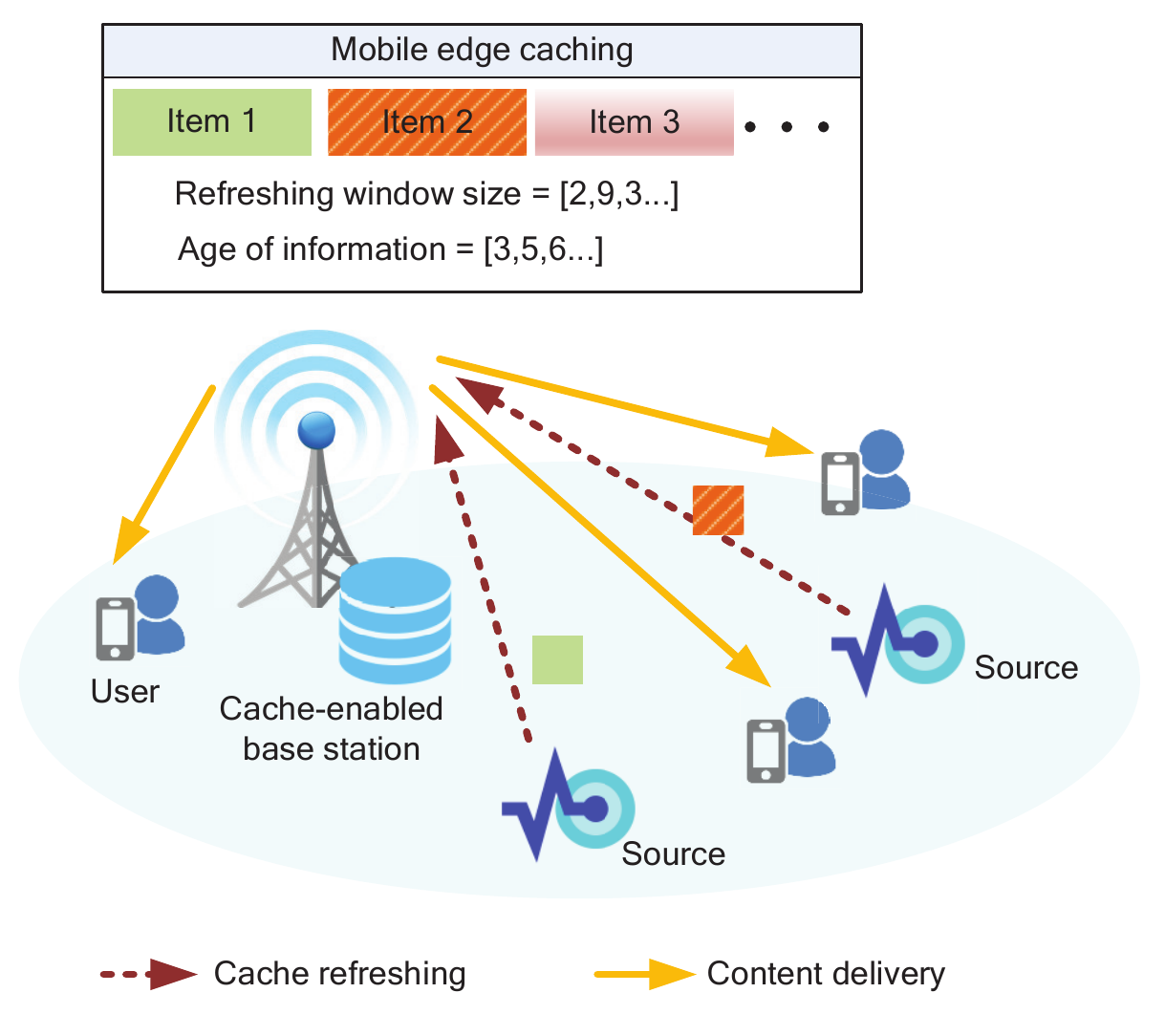}
    	\caption{Mobile edge caching with content refreshing.}
    	\label{fig_scenario}
    \end{figure}
    
    
    \subsection{Traffic Model}
    We consider a typical mobile edge caching network covered by one BS, as shown in Fig.~\ref{fig_scenario}.
    The randomly distributed source nodes monitor the surrounding environment and generate content items, such as the status of traffic jams, the arrival time of a bus, surveillance of a road crossing, availability of parking lots, and store promotions.
    Denote by $\mathcal{C}=\{1,2,3,\cdots,C\}$ the set of generated content items, where $C=|\mathcal{C}|$.  
    Each source node keeps generating new versions of corresponding content items which can reflect the up-to-date information.
    The BS collects the generated content items through wireless transmission, and delivers the cached items to mobile users on demand.
    Mobile users raise requests randomly following Poisson process of arrival rate $\Lambda$.
    Denote by $q_c$ the probability that item-$c$ is requested, where $\sum_{c=1}^{C} q_c = 1$. 
    Accordingly, the requests of item-$c$ also follow Poisson process of rate $\lambda_c \triangleq \Lambda q_c$.

    \subsection{Wireless Transmission Model}
    Denote by $R$ the coverage radius, and $P_\mathrm{BS}$ the transmit power of the BS.
    Suppose the source nodes and users are both uniformly distributed within the coverage of BS.
    Thus, the average service rate of content delivery is given by
    \begin{equation}
    \mu_\mathrm{D} = \underset{r}{\mathds{E}} \left[\frac{B}{L} \log_2 \left(1+\frac{P_\mathrm{BS} r^{-\alpha}}{\sigma^2}\right)\right],
    \end{equation}
    where $r$ is a random variable denoting the distance between a typical mobile user and the BS, $B$ is the available bandwidth, $L$ is the size of a content, $\alpha$ is the path loss exponent of wireless transmission, and $\sigma^2$ is the Gaussian noise.
    As the mobile users are uniformly distributed, the probability distribution of $r$ is $f_r = 2r/R^2$.
    Considering that the transmit power is usually much larger than noise in practice, the average service rate $\mu_\mathrm{D}$ can be approximated as
    \begin{equation}
    \label{eq_mu_d}
    \begin{split}
    & \mu_\mathrm{D} \approx \frac{B}{L} \int_{0}^{R} \left[\log_2 \frac{P_\mathrm{BS}}{\sigma^2} - \alpha \log_2 r \right] \frac{2r}{R^2} \mbox{d} r, \\
    & = \frac{B}{L \ln2} \left[\ln\frac{P_\mathrm{BS}}{\sigma^2} -  \frac{\alpha}{2R^2} \int_{0}^{R^2} \ln x  \mbox{d} x \right] \\
    & = \frac{B}{L} \log_2 \frac{P_\mathrm{BS}\left(R/\sqrt{e}\right)^{-\alpha}}{\sigma^2}.
    \end{split}
    \end{equation}
    Similarly, we obtain the average service rate for content refreshing:
    \begin{equation}
    \label{eq_mu_u}
    \begin{split}
    \mu_\mathrm{R} & = \underset{\{r\}}{\mathds{E}} \left[\frac{B}{L} \log_2 \left(1+\frac{P_\mathrm{Source} r^{-\alpha}}{\sigma^2}\right)\right],\\
    & \approx \frac{B}{L} \log_2 \frac{P_\mathrm{Source}\left(R/\sqrt{e}\right)^{-\alpha}}{\sigma^2},
    \end{split}
    \end{equation}
    where $P_\mathrm{Source}$ is the transmit power of source nodes.
    Notice that Eqs.~(\ref{eq_mu_d}) and (\ref{eq_mu_u}) are both conservative approximations.
    
    \begin{figure}[!t]
    	\centering
    	\includegraphics[width=2.5in]{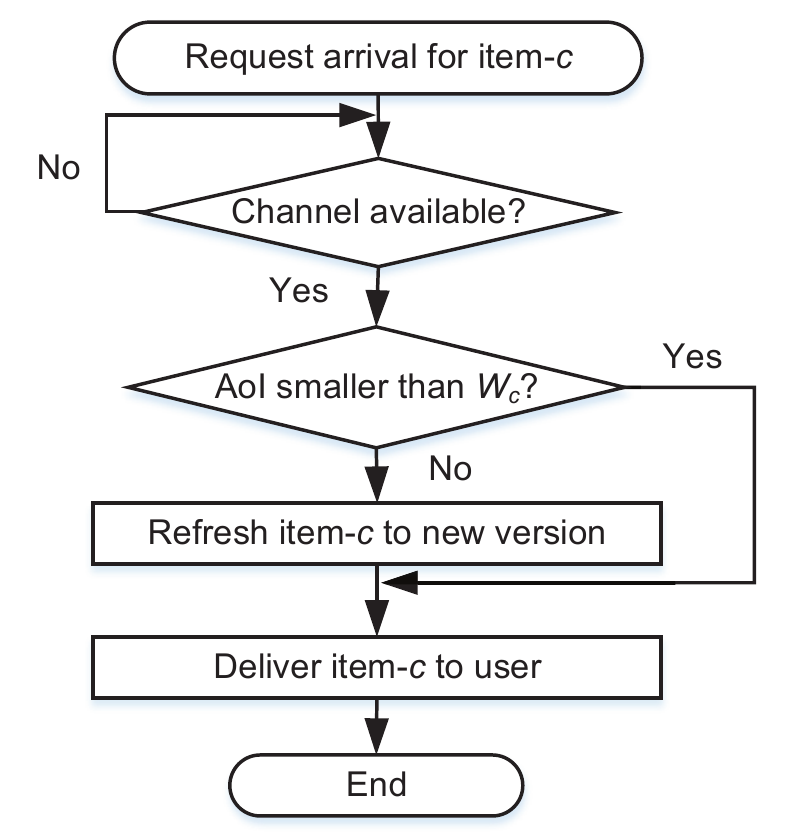}
    	\caption{Freshness-aware refreshing in mobile edge caching.}
    	\label{fig_service_process}
    \end{figure}
    
    \subsection{Service Model}
    Equipped with cache instances, the BSs can store the popular content items and serve mobile users directly through one-hop transmission.
    As such, the BS can save more bandwidth for content delivery with reduced remote content fetching.
    Meanwhile, the freshness-aware cache refreshing scheme is adopted at the BS.
    Specifically, a refreshing window is set for each item to suggest whether the content is fresh or not.
    The BS provides service via the single channel in a First-In-First-Out (FIFO) manner, and the service process is illustrated in Fig.~\ref{fig_service_process}.
    When a user raises a request for item-$c$, it will be served immediately if the channel is not occupied, and wait in a queue otherwise.
    In addition, the BS will always check the freshness before delivering the item.
    If the AoI of item-$c$ is smaller than a threshold $W_{c}$ (defined as the \textit{refreshing window)}, the BS will deliver it to the user directly.
    Otherwise, the BS will fetch the latest version of $c$ from the source node, refresh the cache and then deliver it to users.
    
    The key design issue of the proposed scheme is to set appropriate refreshing windows. 
    On the one hand, increasing the window size will reduce the refreshing frequency, degrading content freshness.
    On the other hand, decreasing the window size will introduce more transmissions due to the frequent content refreshing, degrading the service delay. 
    Therefore, the refreshing window should be optimized to balance the content freshness and service delay.
    However, the analysis of AoI and delay is challenging due to the coupling effect of multi-content refreshing and request arrival, multi-dimensional randomness of traffic arrival and wireless transmissions.
    To address these issues, we first conduct performance analysis for the case of single source, and then extend the result to multi-source cases.
    The M/M/1 queue is adopted to analyze the service process in an approximated manner, which is helpful to reveal the interplay between AoI and delay with respect to the refreshing window size.
    
\section{Single-Source Refreshing Analysis}
    \label{sec_single_source}
    To start with, we analyze the AoI and delay performances of the proposed scheme when all mobile users request the same item from one source node.
    
    \subsection{Queueing Model}
    
    The BS service process can be modeled as a FIFO M/G/1 queue.
    Denote by $T_\mathrm{R}$ and $T_\mathrm{D}$ the time consumed to refresh and deliver a content item, respectively, which are modeled as random variables following exponential distributions due to path loss and channel fading.
    Notice $\mathds{E}[T_\mathrm{R}] = 1/\mu_\mathrm{R}$ and $\mathds{E}[T_\mathrm{D}] = 1/\mu_\mathrm{D}$. 
    The service time of a request is given by
    \begin{equation}
    X=\left\{ \begin{array}{ll} 
    T_\mathrm{D}, & I=0,\\
    T_\mathrm{D} + T_\mathrm{R}, & I = 1,
    \end{array} \right.
    \end{equation}
    where $I$ is a zero-one indicator showing whether the request is served with or without cache refreshing.
    Denote by $p$ the probability that $I=1$.
    The mean and variance of the service time distribution are given by
    \begin{equation}
    \label{eq_X_mean}
    \mathds{E} [X] = \frac{p}{\mu_\mathrm{R}} + \frac{1}{\mu_\mathrm{D}},
    \end{equation}
    and
    \begin{equation}
    \label{eq_coeff_X}
    \begin{split}
    & var[X] = \mathds{E}[X^2] - \mathds{E}^2[X]\\
    & = p \mathds{E} \left[ \left(T_\mathrm{R}+T_\mathrm{D}\right)^2 \right] + (1-p) \mathds{E}[T_\mathrm{D}^2] -\mathds{E}^2[X]\\
    & = \frac{1}{\mu_\mathrm{D}^2} + \frac{1}{\mu_\mathrm{R}^2} - \frac{(1-p)^2}{\mu_\mathrm{u}^2}.
    \end{split}
    \end{equation}
    Notice that $\mathds{E} [X]$ and $var[X]$ both depend on $p$.
    Thus, the refreshing probability should be derived to analyze the service delay and AoI.
    However, $p$ depends on both the arrival and departure processes of the M/G/1 queue, which cannot be derived.
    To get insights, we introduce the M/M/1 queue approximation, where the customer arrival and service rates are set to $\Lambda$ and $1/\mathds{E}[X]$, respectively.
    According to (\ref{eq_coeff_X}), $var[X] \leq \mathds{E}^2[X]$ if $p>1-\frac{\mu_\mathrm{R}}{\mu_\mathrm{D}}$, and vice versa.
    Thus, the M/M/1 approximation is conservative regarding the average delay when the refreshing probability is larger than a certain threshold, according to the Pollaczek--Khinchine formula.
    Specifically, the approximation is always conservative if $\mu_\mathrm{R} \geq \mu_\mathrm{D}$.
    
    \subsection{Content Refreshing Probability}
    
    \begin{figure}[!t]
    	\centering
    	\includegraphics[width=4in]{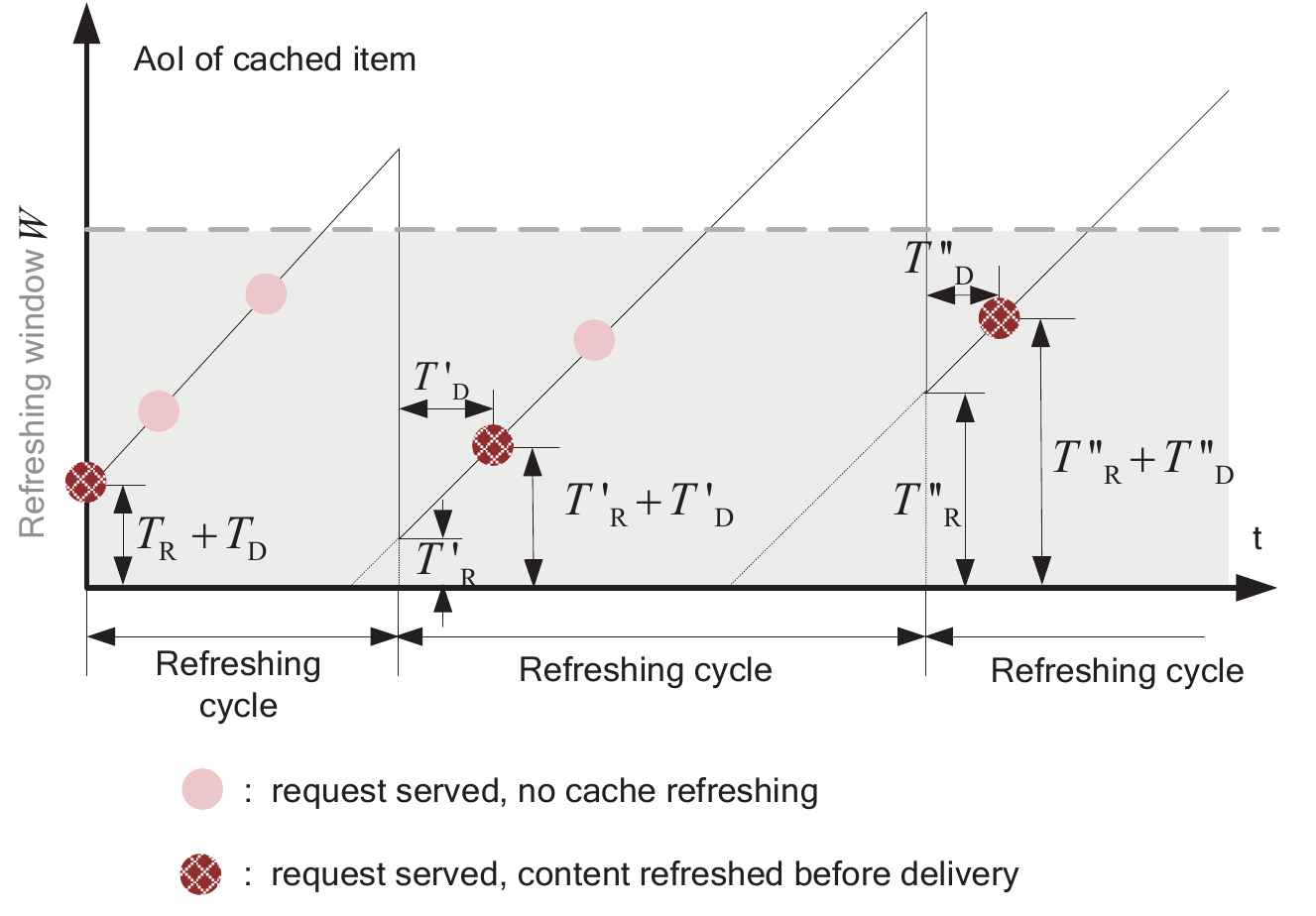}
    	\caption{AoI variation with content refreshing, single-source.}
    	\label{fig_refresh_single}
    \end{figure}

    The content refreshing process depends on the AoI of cached item, as illustrated in Fig.~\ref{fig_refresh_single}.
    The solid line depicts the AoI of the cached item at the BS, and each circle denotes that a user is served with requested content.
    In addition, the solid circles denote the requests directly served without cache refreshing, while the shadowed ones represent the requests which trigger refreshing.
    Suppose there is one request served at the initial time with refreshing.
    Accordingly, the AoI is $T_\mathrm{R}+T_\mathrm{D}$, i.e., the overall time consumed in content fetching and delivery.
    Then, the AoI increases with time, until another request triggers cache refreshing.
    As the cached items are refreshed only when the AoI exceeds the refreshing window $W$, the requests served during time $[0, W-T_\mathrm{R}-T_\mathrm{D}]$ will not trigger refreshing.
    Then, the content refreshing probability can be derived according to queuing theory, given by Theorem~1.
    
    \textbf{Theorem~1.} For the single-source case, a request triggers cache refreshing with probability
    \begin{equation}
    \label{eq_p_W_lambda}
    p = \frac{1}{(W-\frac{1}{\mu_\mathrm{R}}-\frac{1}{\mu_\mathrm{D}})\Lambda} \left[1-e^{-(W-\frac{1}{\mu_\mathrm{R}}-\frac{1}{\mu_\mathrm{D}})\Lambda}\right],
    \end{equation} 
    where $W$ is the cache refreshing window size, $\Lambda$ is the request arrival rate, $\mu_\mathrm{R}$ and $\mu_\mathrm{D}$ are the average service rate of content fetching and delivery, respectively.
    In addition, the refreshing frequency is given by
    \begin{equation}
    p \Lambda = \frac{1}{W-\frac{1}{\mu_\mathrm{R}}-\frac{1}{\mu_\mathrm{D}}} \left[1-e^{-(W-\frac{1}{\mu_\mathrm{R}}-\frac{1}{\mu_\mathrm{D}})\Lambda}\right].
    \end{equation}

    \emph{Proof.} For a stable M/M/1 queueing system, the customer departure process also follows Poisson process of rate $\Lambda$.
    Accordingly, the inter-departure time between two requests follows exponential distribution.
    Define the period between two successive refreshing operations as a refreshing cycle (as shown in Fig.~\ref{fig_refresh_single}), wherein $N$ requests are served directly without cache refreshing.
    The length of a refreshing cycle is random and depends on the arrival/departure of requests which trigger cache refreshing.
    However, the $N$ directly served requests should all happen within the refreshing window under the proposed scheme.
    Accordingly, $N$ follows Poisson distribution of mean $(W-T_\mathrm{R}-T_\mathrm{D})\Lambda$.
    Thus, the refreshing probability can be obtained:
    \begin{equation}
    \begin{split}
    p & =\mathds{E}\left[\frac{1}{N+1}\right] = \sum_{n=0}^{\infty} \frac{1}{n+1} \frac{\bar{N}^{n}}{n!} e^{-\bar{N}}\\
    & = \frac{1}{\bar{N}} \sum_{n=1}^{\infty} \frac{\bar{N}^n}{n!} e^{-\bar{N}} = \frac{1-e^{-\bar{N}}}{\bar{N}},
    \end{split}
    \end{equation}
    where $\bar{N} = (W-\frac{1}{\mu_\mathrm{R}}-\frac{1}{\mu_\mathrm{D}})\Lambda$, denoting the average number of requests directly served per refreshing cycle.  
    Theorem~1 is thus proved. \hfill \rule{4pt}{8pt}\\
    
    Take the first- and second-order derivatives of $p$ with $\bar{N}$:
    \begin{equation}
    \frac{\partial p}{\partial \bar{N}} = - \frac{1-e^{-\bar{N}}}{\bar{N}^2} +\frac{e^{-\bar{N}}}{\bar{N}} = \frac{e^{-\bar{N}}}{\bar{N}^2}\left( \bar{N} +1-e^{\bar{N}} \right),
    \end{equation}
    and
    \begin{equation}
    \frac{\partial^2 p}{\partial \bar{N}^2} = \frac{2e^{-\bar{N}}}{\bar{N}^3} \left( e^{\bar{N}} -1-\bar{N} -\frac{\bar{N}^2}{2} \right).
    \end{equation}
    By taking Taylor's series of $e^x$, we can prove $e^x\geq 1+x+\frac{1}{2}x^2$, $\forall x\geq 0$.
    Therefore, $\frac{\partial p}{\partial \bar{N}} \leq 0$ and $\frac{\partial^2 p}{\partial \bar{N}^2} \geq 0$.
    Thus, $p$ is a convexly decreasing function with respect to $\bar{N}$.
    As $\bar{N}$ is a linear increasing function of $\Lambda$ and $W$, the refreshing probability also shows convexity with respect to $\Lambda$ and $W$.
    In the same way, we can prove that the content refreshing frequency $p \Lambda$ is a concave increasing function with respect to $\Lambda$ and a convexly decreasing function with respect to $W$.
    
    The important insight of Theorem~1 is that the proposed scheme restrains the BS from frequently refreshing one item of high request rate (i.e., popular content items), since the refreshing probability $p$ decreases with $\Lambda$. 
    However, the popular items are still refreshed more frequently under the same refreshing window size $W$. 
    Furthermore, the refreshing probability indicates whether the item is worth to cache or not. 
    Notice that $p\rightarrow 1$ as the $\Lambda \rightarrow 0$ or $W \rightarrow \frac{1}{\mu_\mathrm{R}}+\frac{1}{\mu_\mathrm{D}}$.
    Therefore, the unpopular items with strict freshness requirements always trigger refreshing, and have no need to cache if the storage resource is constrained.

    \subsection{Age of Information Analysis}
    
    \textbf{Theorem~2.} For the single-source case, the average AoI of user-received contents is given by
    \begin{equation}
    \bar{A} = \frac{1}{2} \left(W+\frac{1}{\mu_\mathrm{R}}+\frac{1}{\mu_\mathrm{D}}\right) - \frac{1}{2\Lambda} \left( 1-e^{-\left(W-\frac{1}{\mu_\mathrm{R}}-\frac{1}{\mu_\mathrm{D}}\right)\Lambda} \right),
    \end{equation} 
    under the proposed freshness-aware refresh scheme.
    
    \emph{Proof.} Consider one refresh cycle wherein $N+1$ requests are served.
    The first request triggers cache refresh, and the AoI of user received content equals to $T_\mathrm{R}+T_\mathrm{D}$.
    The other $N$ requests are served during $[T_\mathrm{R}+T_\mathrm{D}, W]$.
    As the departure process of a M/M/1 queue follows Poisson process, the $N$ requests distribute uniformly in $[T_\mathrm{R}+T_\mathrm{D}, W]$.
    Therefore, the average AoI of contents received by mobile users is given by
    \begin{equation}
    \begin{split}
    \bar{A} & = \underset{\{N,T_\mathrm{R},T_\mathrm{D}\}}{\mathds{E}} \left[ \frac{1}{N+1} \left( T_\mathrm{R} +T_\mathrm{D} + N\frac{T_\mathrm{R}+T_\mathrm{D}+W}{2}\right)\right]\\
    & = \underset{\{N,T_\mathrm{R},T_\mathrm{D}\}}{\mathds{E}} \left[  \frac{W+T_\mathrm{R}+T_\mathrm{D}}{2} - \frac{W-T_\mathrm{R}-T_\mathrm{D}}{2(N+1)} \right]\\
    & = \frac{1}{2} \!\left(\!W\!+\!\frac{1}{\mu_\mathrm{R}}\!+\!\frac{1}{\mu_\mathrm{D}}\!\right) \!-\! \frac{1}{2} \!\left(\! W\!-\!\frac{1}{\mu_\mathrm{R}}\!-\!\frac{1}{\mu_\mathrm{D}}\!\right)\!\underset{\{N\}}{\mathds{E}}\left[\!\frac{1}{N+1}\!\right] \\
    & = \frac{1}{2} \left(W+\frac{1}{\mu_\mathrm{R}}+\frac{1}{\mu_\mathrm{D}}\right) - \frac{1}{2\Lambda} \left( 1-e^{-\left(W-\frac{1}{\mu_\mathrm{R}}-\frac{1}{\mu_\mathrm{D}}\right)\Lambda} \right),
    \end{split}
    \end{equation}
    Theorem~2 is thus proved. \hfill \rule{4pt}{8pt}\\
    
    Based on Theorem~2, we further analyze the relationship between AoI and refreshing window size.
    Take the first- and second-order derivatives of $\bar{A}$ with respect to $W$:
    \begin{equation}
    \begin{split}
    \label{eq_A_Win_first}
    \frac{\partial \bar{A}}{\partial W} & = \frac{1}{2} - \frac{1}{2\Lambda} e^{ -\left(W-\frac{1}{\mu_\mathrm{R}}-\frac{1}{\mu_\mathrm{D}}\right)\Lambda } \Lambda\\
    & = \frac{1}{2} \left[ 1- e^{ -\left(W-\frac{1}{\mu_\mathrm{R}}-\frac{1}{\mu_\mathrm{D}}\right)\Lambda } \right] \geq 0,
    \end{split}
    \end{equation}
    
    \begin{equation}
    \label{eq_A_Win_second}
    \frac{\partial^2 \bar{A}}{\partial W^2} = \Lambda e^{ -\left(W-\frac{1}{\mu_\mathrm{R}}-\frac{1}{\mu_\mathrm{D}}\right)\Lambda } \geq 0.
    \end{equation}
    Therefore, $A$ is a convexly increasing function with respect to $W$. 
    In specific, $\frac{\partial^2 \bar{A}}{\partial W^2} \rightarrow 0$ and $\frac{\partial \bar{A}}{\partial W} \rightarrow \frac{1}{2}$ as $W\rightarrow \infty$, which means $\bar{A}$ is an asymptotically linear function of $W$.
    
    Theorem~2 also indicates how the request arrival rate influences the content freshness.
    Take the first- and second-order derivatives of $\bar{A}$ with respect to $\Lambda$:
    \begin{equation}
    \label{eq_A_lambda_1}
    \begin{split}
    \frac{\partial \bar{A}}{\Lambda} &  = \frac{\partial \bar{A}}{\partial \bar{N}} \frac{\partial \bar{N}}{\partial \bar{\Lambda}} = - \frac{1}{2} \left[ -\frac{1}{\bar{N}^2 } \left( 1-e^{-\bar{N}}\right) +\frac{1}{\bar{N}}e^{-\bar{N}} \right]\\
    & = -\frac{e^{-\bar{N}}}{2\bar{N}^2} \left[ 1+\bar{N} -e^{\bar{N}} \right] \geq 0,
    \end{split}
    \end{equation}
    \begin{equation}
    \label{eq_A_lambda_2}
    \begin{split}
    \frac{\partial^2 \bar{A}}{\partial \Lambda^2} 
    \!&\! = \!-\! \left( \frac{e^{\!-\!\bar{N}}}{\bar{N}^3} \!+\! \frac{e^{\!-\!\bar{N}}}{2\bar{N}^2}\right) \! \left(e^{\bar{N}}\!-\!\bar{N}-1\right) \! + \! \frac{e^{\!-\!\bar{N}}}{2\bar{N}^2} \! \left(e^{\bar{N}}\!-\! 1 \! \right)\! \\
    & = -\frac{e^{-\bar{N}}}{\bar{N}^3} \left( e^{\bar{N}} -1-\bar{N} - \frac{\bar{N}^2}{2} \right) \leq 0,
    \end{split}
    \end{equation}
    where $\bar{N}= \Lambda\left(W-\frac{1}{\mu_\mathrm{R}}-\frac{1}{\mu_\mathrm{D}}\right)$.
    Accordingly, the average AoI $\bar{A}$ is a concave increasing function with respect to the request rate.
    Two special cases are provided.		
    
    \emph{\textbf{(1) Unpopular content:} $\Lambda \rightarrow 0$} \\	
    The average AoI and content refreshing probability are given by
    \begin{equation}
    \label{eq_A_lambda_3}
    \begin{split}
    \lim\limits_{\Lambda \rightarrow 0} \bar{A} & = \frac{1}{2} \left(W+\frac{1}{\mu_\mathrm{R}}+\frac{1}{\mu_\mathrm{D}}\right) - \frac{1}{2} \left(W-\frac{1}{\mu_\mathrm{R}}-\frac{1}{\mu_\mathrm{D}}\right)\\
    & = \frac{1}{\mu_\mathrm{R}} +\frac{1}{\mu_\mathrm{D}},
    \end{split}
    \end{equation}
    and 
    \begin{equation}
    \lim\limits_{\Lambda \rightarrow 0} p = \lim\limits_{\bar{N} \rightarrow 0} \frac{1-e^{-\bar{N}}}{\bar{N}} = 1.
    \end{equation}
    In this case, all requests are served with cache refreshing, and the mobile edge caching is not utilized. 
    
    \emph{\textbf{(2) Popular content:} $\Lambda \rightarrow \infty$}\\	
    The average AoI and content refreshing probability are given by
    \begin{equation}
    \lim\limits_{\Lambda \rightarrow \infty} \bar{A} = \frac{1}{2} \left(W+\frac{1}{\mu_\mathrm{R}}+\frac{1}{\mu_\mathrm{D}}\right),
    \end{equation}
    and 
    \begin{equation}
    \label{eq_A_lambda_4}
    \lim\limits_{\Lambda \rightarrow \infty} p = \lim\limits_{\bar{N} \rightarrow \infty} \frac{1-e^{-\bar{N}}}{\bar{N}} = 0.
    \end{equation}
    The results indicate that almost all requests are directly served without cache refreshing. In this case, the mobile edge caching plays a key role.

    \subsection{Service Delay Analysis}
    
    \textbf{Theorem~3.} For the single-source case, the average service delay of the proposed freshness-aware cache refreshing scheme is given by
    \begin{equation}
    \label{eq_delay_mean_single}
    \bar{D} =  \frac{\frac{p}{\mu_\mathrm{R}}+\frac{1}{\mu_\mathrm{D}}}{1-\frac{p\Lambda}{\mu_\mathrm{R}}-\frac{\Lambda}{\mu_\mathrm{D}}},
    \end{equation} 
    where $p$ is the refreshing probability given by Eq.~(\ref{eq_p_W_lambda}).\\
    
    \emph{Proof.} According to the M/M/1 queueing model, the average service delay is given by $\bar{D} = \frac{1}{\frac{1}{\mathds{E}[X]}-\Lambda}$. Substituting Eq.~(\ref{eq_X_mean}), Theorem~3 can be proved. \hfill \rule{4pt}{8pt}\\
    
    According to (\ref{eq_delay_mean_single}), the average delay increases with the refreshing probability $p$.
    This is reasonable since the queueing delay increases with traffic load.
    As $p$ decreases with $W$, the average delay decreases with the refreshing window. 
    On the other hand, the average AoI increases with $W$ according to Theorem~2.
    Thus, the average AoI and service delay can tradeoff by adjusting the refreshing window size.	
    Notice that
    \begin{equation}
    \label{eq_D_max}
    \lim\limits_{W\rightarrow\frac{1}{\mu_\mathrm{R}} +\frac{1}{\mu_\mathrm{D}}} \bar{D} = \lim\limits_{p \rightarrow 1} \bar{D} = \frac{1}{\frac{\mu_\mathrm{R}\mu_\mathrm{D}}{\mu_\mathrm{R}+\mu_\mathrm{D}}-\Lambda} \triangleq \bar{D}_\mathrm{max},
    \end{equation}
    \begin{equation}
    \label{eq_D_min}
    \lim\limits_{W\rightarrow\infty} \bar{D} = \lim\limits_{p \rightarrow \infty} \bar{D} = \frac{1}{\mu_\mathrm{D}-\Lambda} \triangleq \bar{D}_\mathrm{min},
    \end{equation}
    indicating how much delay can be traded by sacrificing the content freshness.
    Examples are illustrated to offer insights.
    
    Case-1: $\mu_\mathrm{R}=\mu_\mathrm{D}$, $\Lambda = \frac{1}{2} \frac{1}{\frac{1}{\mu_\mathrm{R}}+\frac{1}{\mu_\mathrm{D}}}$: $\bar{D}_\mathrm{min}=\frac{1}{3}\bar{D}_\mathrm{max}$;
    
    Case-2: $\mu_\mathrm{R}=\mu_\mathrm{D}$, $\Lambda = \frac{9}{10} \frac{1}{\frac{1}{\mu_\mathrm{R}}+\frac{1}{\mu_\mathrm{D}}}$: $\bar{D}_\mathrm{min}=\frac{1}{11}\bar{D}_\mathrm{max}$;
    
    Case-3: $\mu_\mathrm{R}=\frac{1}{2}\mu_\mathrm{D}$, $\Lambda = \frac{1}{2} \frac{1}{\frac{1}{\mu_\mathrm{R}}+\frac{1}{\mu_\mathrm{D}}}$: $\bar{D}_\mathrm{min}=\frac{1}{5}\bar{D}_\mathrm{max}$.
    
    Case-1 corresponds to the case that content fetching and delivery share the same transmission rates, and the system is half-loaded without mobile caching.
    Case-2 is more heavily loaded, but has the same channel condition of Case-1.
    Case-3 is half-loaded, whereas the file fetching takes longer time, corresponding to the case that source nodes use lower transmit power.
    These cases suggest that the proposed scheme can effectively reduce the average delay by restraining frequent cache refreshing.
    For example, the delay can be reduced by 60\% to 90\% according to the three cases.
    In addition, increasing refreshing window size is more beneficial on delay reduction in heavily-loaded networks with lower transmission rate of source nodes.
    
\section{Multi-Source Refreshing Optimization}
	\label{sec_multi_source}
	\begin{figure}[!t]
		\centering
		\includegraphics[width=4in]{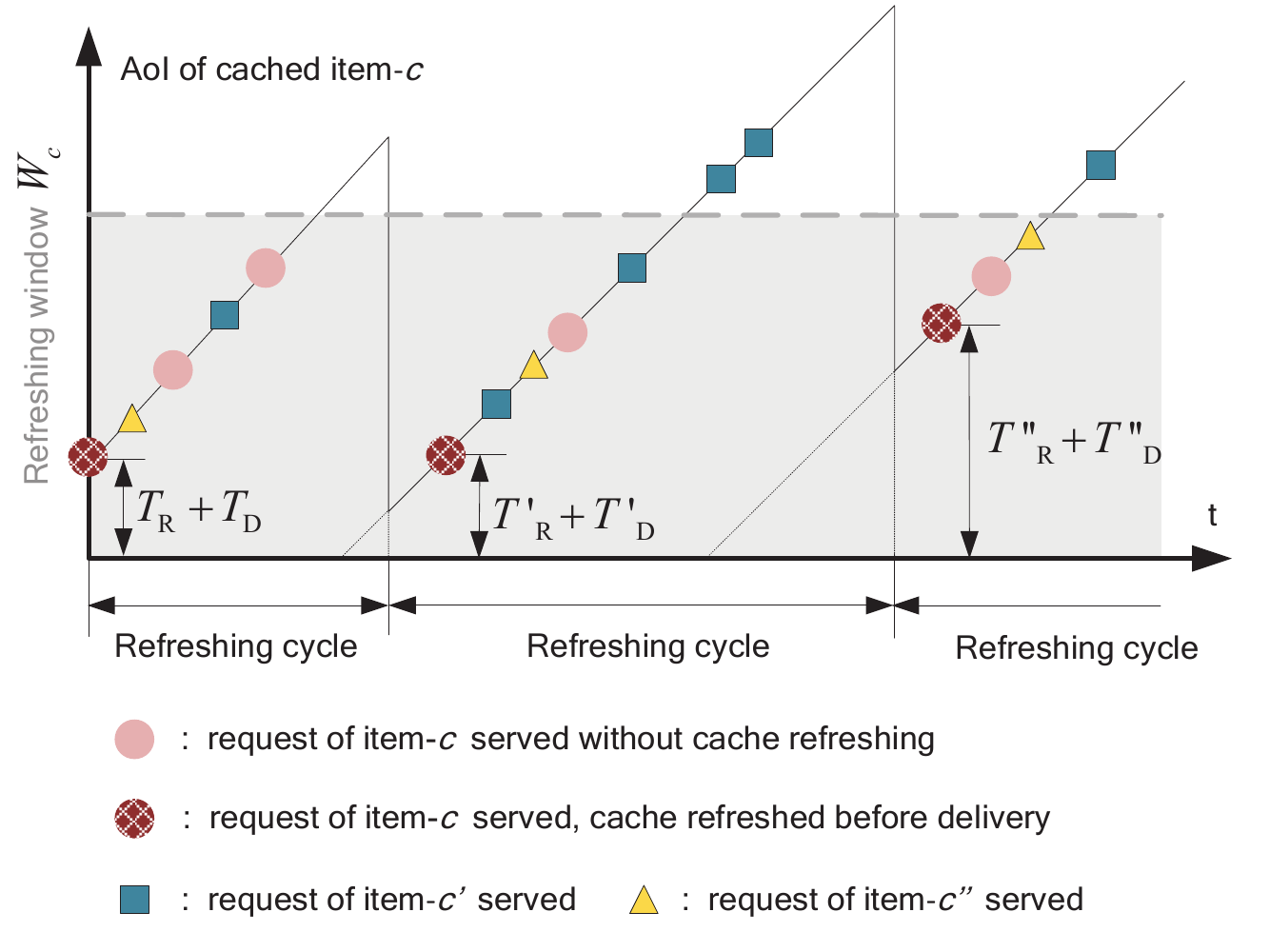}
		\caption{AoI variation with content refreshing, multi-source.}
		\label{fig_refresh_multi}
	\end{figure}
	
	In this section, the analytical results are further extended to the multi-source scenario, whereby the refreshing window size is optimized to minimize the average delay under the average AoI constraint of all sources.
	
	\subsection{AoI and Delay Analysis}
	
	The queueing model can be extended to depict the service process of multi-source case, whereas the request arrival and service rates are different.
	Similarly, we study the departure process of the queue to analyze the refreshing probability of each item, whereby the average service rate can be derived.
	Then, the average AoI and delay can be obtained, given by Theorem~4.
	
	\textbf{Theorem~4.} In the multi-source scenario, the average AoI of user received content item-$c$ is given by
	\begin{equation}
	\label{eq_AoI_multi}
	\bar{A}_{\mathrm{M},c} \!= \!\frac{1}{2} \!\left[ \!W_c \!+\! \frac{1}{\mu_\mathrm{R}} \!+\! \frac{1}{\mu_\mathrm{D}} \!\right]\! - \! \frac{1}{2\lambda_c} \! \left[ \!1\!-\! e^{-\lambda_c\left(W_c\!-\!\frac{1}{\mu_\mathrm{R}}\!-\!\frac{1}{\mu_\mathrm{D}}\right)}\!\right]\!.
	\end{equation}
	The average service delay is given by
	\begin{equation}
	\label{eq_D_multi}
	\bar{D}_\mathrm{M} = \frac{\frac{\bar{P}}{\mu_\mathrm{R}}+\frac{1}{\mu_\mathrm{D}}}{1-\frac{\bar{P}\Lambda}{\mu_\mathrm{R}}-\frac{\Lambda}{\mu_\mathrm{D}}},
	\end{equation}
	where $\Lambda=\sum_{c=1}^{C} \lambda_c$ is the total request arrival rate of all items, $\bar{P}$ is the average cache refreshing probability
	\begin{equation}
	\label{eq_P_aver_multi}
	\bar{P} = \frac{\sum_{c=1}^{C}\lambda_c p_c}{\sum_{c=1}^{C}\lambda_c},
	\end{equation}
	and $p_c$ is the cache refreshing probability of item-$c$
	\begin{equation}
	\label{eq_p_c}
	p_c = \frac{1-e^{-\lambda_c \left(W_c-\frac{1}{\mu_\mathrm{R}}-\frac{1}{\mu_\mathrm{D}}\right)}}{\lambda_c \left(W_c-\frac{1}{\mu_\mathrm{R}}-\frac{1}{\mu_\mathrm{D}}\right)}.
	\end{equation}
	
	\emph{Proof.} The BS service process is approximated as a FIFO M/M/1 queue with arrival rate of $\Lambda$.
	If the queue is stable, the departure process of all requests follows Poisson process, as illustrated in Fig.~\ref{fig_refresh_multi}.
	As for item-$c$, the departure process also follows Poisson process of rate $\lambda_c$, as a randomly thinning of all requests.
	Therefore, the refreshing probability can be derived in the same way as the single-source scenario, given by (\ref{eq_p_c}).
	The average refreshing probability of all items can be obtained as (\ref{eq_P_aver_multi}), and the average service time is given by
	\begin{equation}
	\mathds{E}[X_\mathrm{M}] = \frac{\bar{P}}{\mu_\mathrm{R}} +\frac{1}{\mu_\mathrm{D}}.
	\end{equation}
	Thus, the average delay is given by $\bar{D}_\mathrm{M} = \frac{1}{\frac{1}{\mathds{E}[X_\mathrm{M}]}-\Lambda}$, which can be rewritten as (\ref{eq_D_multi}).
	Notice that the average AoI of item-$c$ is not influenced by the service of other items if the queue is stable, and shares the same form of the single-source case as (\ref{eq_AoI_multi}).
	Hence, Theorem~4 is proved. \hfill \rule{4pt}{8pt}\\
	
	Theorem~4 indicates the main influencing factors on the service delay and average AoI of user-received contents.
	When the BS is not overloaded (i.e., stable queue), the average AoI of an item mainly depends on the refreshing window size and the request arrival rate of the corresponding item, and merely influenced by the other items.
	Unlikely, the average delay depends on the intensity and refreshing window of all items, showing a coupling effect.
	This result indicates that the content freshness can be guaranteed by setting appropriate refreshing windows, regardless of the other items.
	Furthermore, the refreshing windows of each item should be jointly optimized to enhance the system-level delay performance.
	
	\subsection{Refreshing Window Optimization}
	
	Based on the result of Theorem~4, we optimize the refreshing window to minimize the average delay while meeting the AoI requirement. 
	The problem can be formulated as follows.
	
	\begin{subequations}
		\label{eq_P2}
		\begin{align}
		\min\limits_{\{W_c\}}~~&~~\bar{D}_\mathrm{M} \\
		\mbox{(P1)}~~~~~s.t.~~&~~ \frac{1}{\Lambda}\sum_{c=1}^{C} \lambda_c \bar{A}_{\mathrm{M},c} \leq \hat{A},\\
		&~~ \sum_{c=1}^{C} \lambda_c < \frac{1}{\frac{\bar{P}}{\mu_\mathrm{R}}+\frac{1}{\mu_\mathrm{D}}},\\
		&~~W_c \geq \frac{1}{\mu_\mathrm{R}} + \frac{1}{\mu_\mathrm{D}}, ~~c=1,2,...,C, 
		\end{align}
	\end{subequations}     
	where $\hat{A}$ is the required average AoI, and constraint (\ref{eq_P2}c) guarantees the system stability. 
	(P1) applies to the case where contents have different popularities but require the same level of freshness (eg., a network slice providing the same type of content service).
	
	Suppose (P1) is feasible and the average delay is not infinity.
	Thus, the system is stable and constraint (\ref{eq_P2}c) holds. 
	Therefore, we focus on the freshness constraint to solve (P1).	
	For notation simplicity, (P1) can be written as 
	\begin{subequations}
		\label{eq_P2_1}
		\begin{align}
		\min\limits_{\{\bar{N}_c\}}~&~\sum_{c=1}^{C} \frac{1-e^{-\bar{N}_c}}{\bar{N}_c} \lambda_c \\
		\mbox{(P2)}~s.t.~&~\sum_{c=1}^{C} \left( \bar{N}_c \!+\!e^{-\bar{N}_c} \right) \leq 2\Lambda\hat{A} \!+\! C \!-\! 2\Lambda \left(\frac{1}{\mu_\mathrm{R}} \!+\!\frac{1}{\mu_\mathrm{D}}\right),\\
		&~~N_c \geq 0, ~~c=1,2,...,C, 
		\end{align}
	\end{subequations} 
	where $\bar{N}_c = \lambda_c \left( W_c- \frac{1}{\mu_\mathrm{R}} - \frac{1}{\mu_\mathrm{D}} \right)$.
	The objective function of (P2) is to minimize the average refreshing probability, which is equivalent to minimize the average delay.
	(P2) is proved to be convex regarding $\bar{N}_c$, and the details are omitted due to page limit.
	Thus, (P2) can be addressed through convex optimization toolboxes.
	
	We further apply the Lagrange method and analyze the optimal condition to offer insights on the setting of refreshing window.
	By taking derivative of the objective function and the constraint with respect to $\bar{N}_c$, the optimal condition of (P2) can be obtained:
	\begin{equation}
	\label{eq_P2'_opt}
	\frac{\bar{N}_c e^{-\bar{N}_c} -1 +e^{-\bar{N}_c}}{\bar{N}_c^2} \lambda_c + \nu_0 \left( 1-e^{-\bar{N}_c} \right) - \nu_c = 0,
	\end{equation}
	where $\nu_0$ and $\nu_c$ are the Lagrange multipliers corresponding to the two constraints.
	Assume $\bar{N}_c\neq 0$ and $\nu_c=0$, and (\ref{eq_P2'_opt}) can be written as
	\begin{equation} 
	\label{eq_P2'_opt_1}
	\nu_0 = \frac{\lambda_c}{\bar{N}_c} \left[ \frac{1}{1-e^{-\bar{N}_c}} -1-\frac{1}{\bar{N}_c} \right].
	\end{equation}		
	Next, we prove that $\nu_0$ increases with $W_c$ and $\lambda_c$.
	Denote by $\tilde{W}_c = W_c - \frac{1}{\mu_\mathrm{R}} - \frac{1}{\mu_\mathrm{D}}$ for notation simplicity.
	Thus, 
	\begin{equation}
	\nu_0 = \frac{1}{\tilde{W}_c} \left[ \frac{1}{1-e^{-\tilde{W}_c\lambda_c}} -\frac{1}{\tilde{W}_c\lambda_c} -1\right].
	\end{equation}
	Take derivative, and we have
	\begin{equation}
	\begin{split}
	& \frac{\partial \nu_0}{\partial \tilde{W}_c} = \frac{\partial \nu_0}{\partial \bar{N}_c} \frac{\partial \bar{N}_c}{\partial \tilde{W}_c} = \frac{\partial }{\partial \bar{N}_c}\left\{ \frac{\lambda_c^2}{\bar{N}_c} \left[ \frac{1}{1-e^{-\bar{N}_c}} -\frac{1}{\bar{N}_c} -1 \right]\right\}\\
	& = \lambda_c^2\left[-\frac{1}{\bar{N}_c^2\left( 1-e^{-\bar{N}_c}\right)} + \frac{e^{-\bar{N}_c}}{\bar{N}_c \left(1-e^{-\bar{N}_c} \right)^2} +\frac{2}{\bar{N}_c^3} +\frac{1}{\bar{N}_c^2}\right]\\
	& = \lambda_c^2\left[\frac{e^{-\bar{N}_c}}{\bar{N}_c\left( 1-e^{-\bar{N}_c}\right)} \left[ \frac{1}{1-e^{-\bar{N}_c}} -\frac{1}{\bar{N}_c} \right] + \frac{2}{\bar{N}_c^3}\right].
	\end{split}
	\end{equation}
	As $1-e^{-x} \leq x,~\forall x\geq 0$, $\frac{\partial \nu_0}{\partial \tilde{W}_c} \geq 0$.
	Accordingly, $\nu_0$ increases with the refreshing window size $W_c$, considering the linear relationship between $W_c$ and $\tilde{W}_c$.
	Take derivative of $\nu_0$ with respect to $\lambda_c$, 
	\begin{equation}
	\begin{split}
	& \frac{\partial \nu_0}{\partial \lambda_c} = \frac{1}{\tilde{W}_c} \left[ - \frac{\tilde{W}_c e^{-\tilde{W}_c \lambda_c}}{\left( 1-e^{-\tilde{W}_c\lambda_c}\right)^2} +\frac{1}{\tilde{W}_c\lambda_c^2} \right] \\
	& = -\frac{e^{-\tilde{W}_c\lambda_c}}{\left( 1-e^{-\tilde{W}_c\lambda_c}\right)^2} + \frac{1}{\left(\tilde{W}_c\lambda_c\right)^2}\\
	& = \left(\frac{1}{\bar{N}_c}+\frac{e^{-\frac{\bar{N}_c}{2}}}{1-e^{-\bar{N}_c}}\right) \left(\frac{1}{\bar{N}_c}-\frac{1}{e^{\frac{\bar{N}}{2}}-e^{-\frac{\bar{N}_c}{2}}}\right).
	\end{split}
	\end{equation}
	Denote by $Z = e^{\frac{\bar{N}}{2}}-e^{-\frac{\bar{N}_c}{2}} - \bar{N}_c$.
	Notice that 
	\begin{equation}
	\frac{\partial Z}{\partial \bar{N}_c} = \frac{1}{2} e^{\frac{\bar{N}}{2}} + \frac{1}{2} e^{-\frac{\bar{N}}{2}} -1 \geq 0.
	\end{equation}
	Thus, $Z|_{\bar{N}_c>0} \geq Z|_{\bar{N}_c=0} = 1-1-0=0$.
	Therefore, $\frac{\partial \nu_0}{\partial \lambda_c} \geq 0$, and $\nu_0$ increases with $\lambda_c$.
	For two items $c$ and $c'$, if $\lambda_c \geq \lambda_{c'}$, the optimal refreshing window size should satisfy $W_c<W_{c'}$ according to the optimal condition (\ref{eq_P2'_opt_1}).
	
	The important insight is that the popular contents should be set with a smaller refreshing window to minimize the service delay while meeting the system-level freshness requirement.
	Although (P1) focuses on the single-class traffic which has the same average AoI requirements, it can be easily extended.
	Specifically, constraint (\ref{eq_P2}b) can be divided into multiple constraints to reflect differentiated freshness requirements.
	The problem is equivalent to minimize the average refreshing probability of each class, which can be decoupled as solved in the same way as (P1).

	\begin{table}[t!]
		\caption{Simulation parameters}
		\label{tab_parameter}
		\centering
		\begin{tabular}{cc}
			\hline
			\hline
			Parameter & Value  \\
			\hline
			coverage radius $R$ & 1000 m \\
			system bandwidth $B$ & 10 MHz \\
			packet size $L$ & 10 KB \\
			path loss factor $\alpha$ & 4  \\
			addictive white noise $\sigma^{2}$ & -95 dbm\\
			transmit power of BS $P_\mathrm{BS}$ & 1 W \\
			transmit power of source nodes $P_\mathrm{Source}$ & 0.1 \\	
			request arrival rate $\Lambda$ & 1000 /s\\ 		
			\hline
			\hline
		\end{tabular}
	\end{table}
	
\section{Simulation and Numerical Results}
    \label{sec_simulation}
    	To validate the theoretical analysis, we conduct system-level simulations on the event-based OMNeT++ simulator, where the user locations, content requests and packet transmission time are generated randomly by the Monte Carlo method.
    	The BS serves the requests in a FIFO manner and implements the proposed cache refreshing scheme.
    	The derived refreshing probability, average AoI and delay are compared with the simulation results for both the single- and multi-source scenarios.	
    	In addition, the refreshing window is optimized in the multi-source scenario, and the influence of important system parameters is illustrated. 
    	Furthermore, the performance of the proposed scheme is also evaluated in a practical urban scenario, by implementing the Veins framework to simulate the real-trace user mobility and traffic demand \cite{Veins}.
    	Simulation parameters are listed in Table~\ref{tab_parameter}, unless stated otherwise.

    	\subsection{Validation of Single-Source Analysis}
    	
\begin{figure}
    		\centering
    		\subfloat[] {\includegraphics[width=3in]{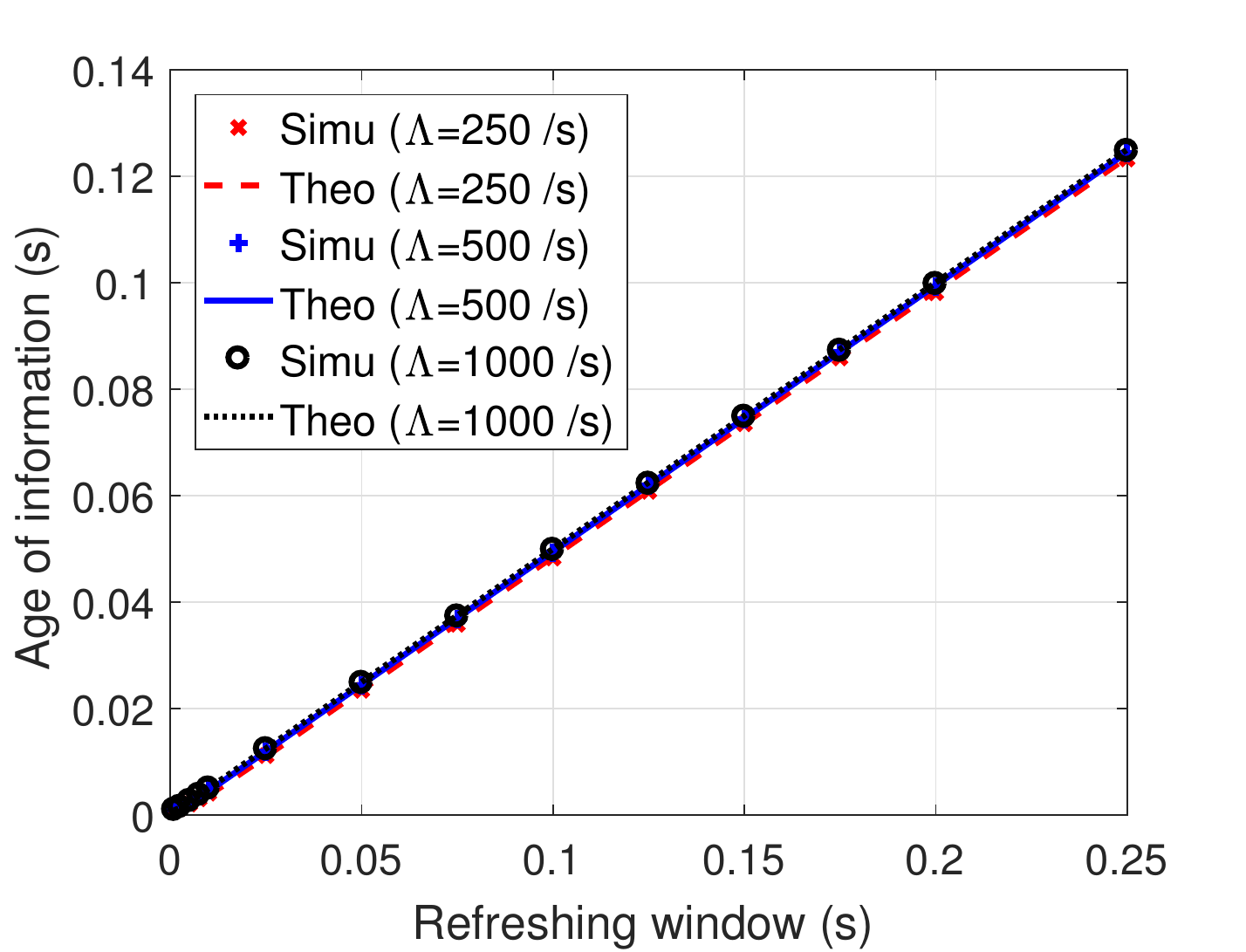}}\\
    		\subfloat[]{\includegraphics[width=3in]{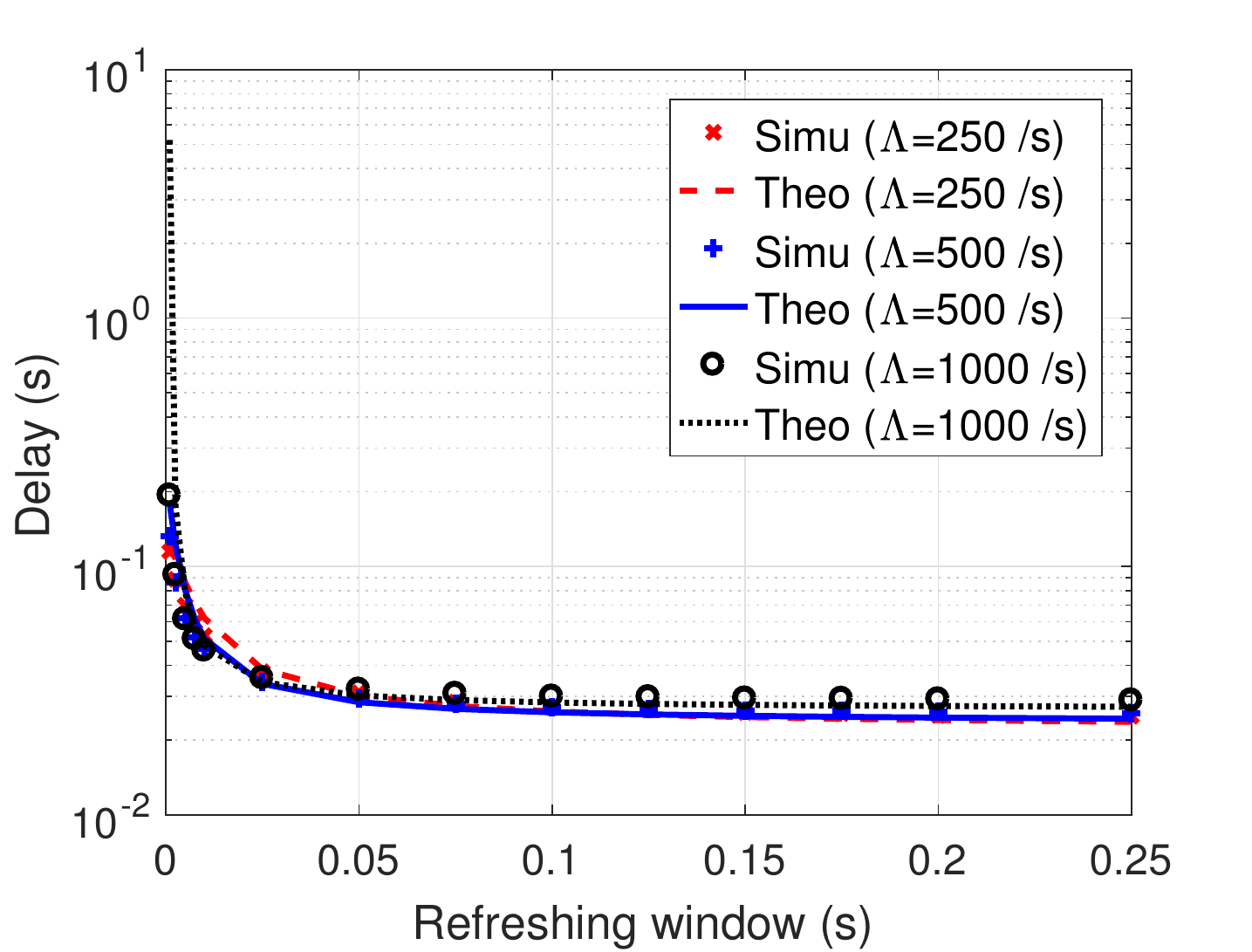}}\\
    		\subfloat[]{\includegraphics[width=3in]{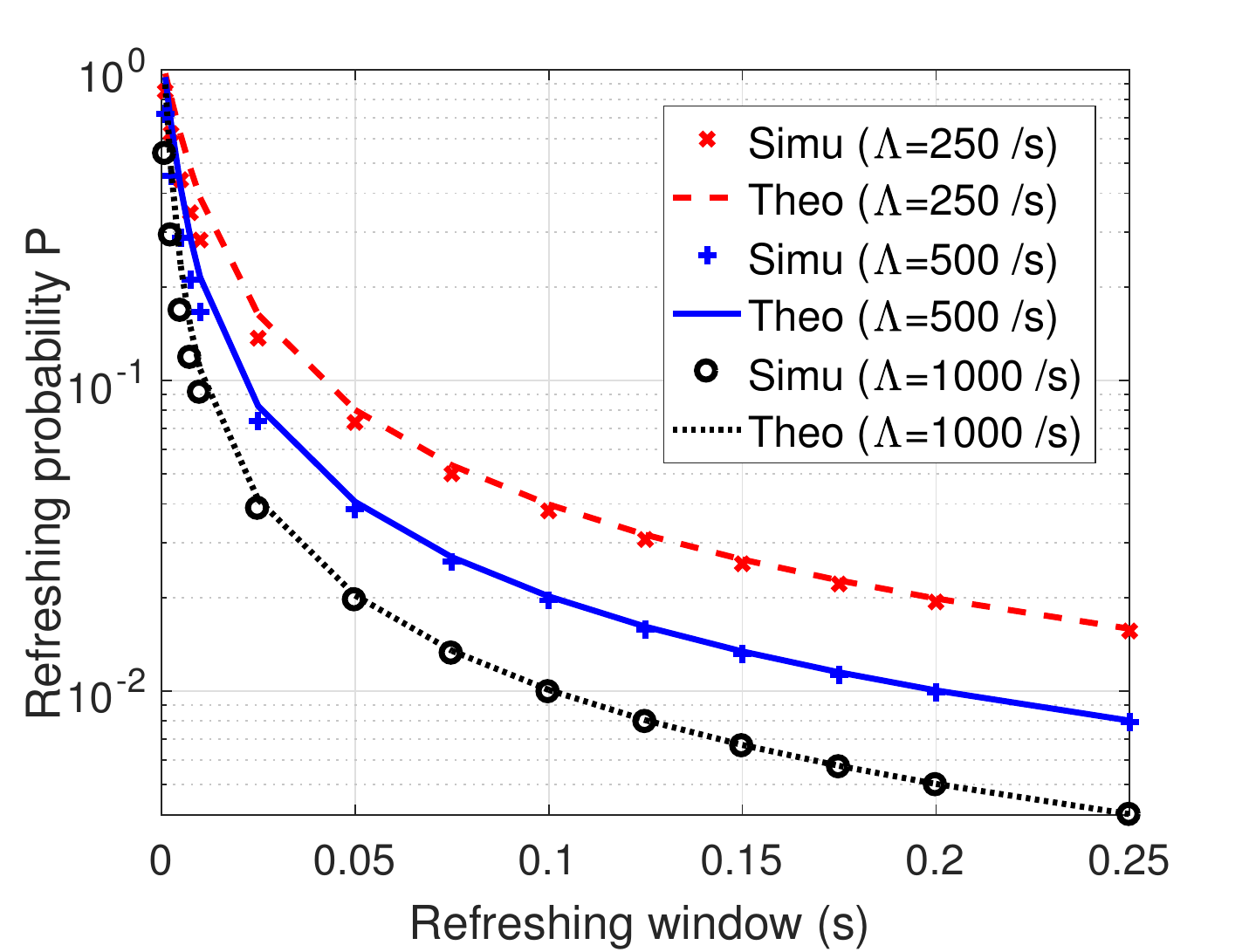}}
    		\caption{Analytical results validation with respect to refreshing window in single-source scenario, (a) Age of information, (b) Delay, (c) Refresh probability.}
    		\label{fig_single_wc}
    	\end{figure}

\begin{figure}
    		\centering
    		\subfloat[] {\includegraphics[width=3in]{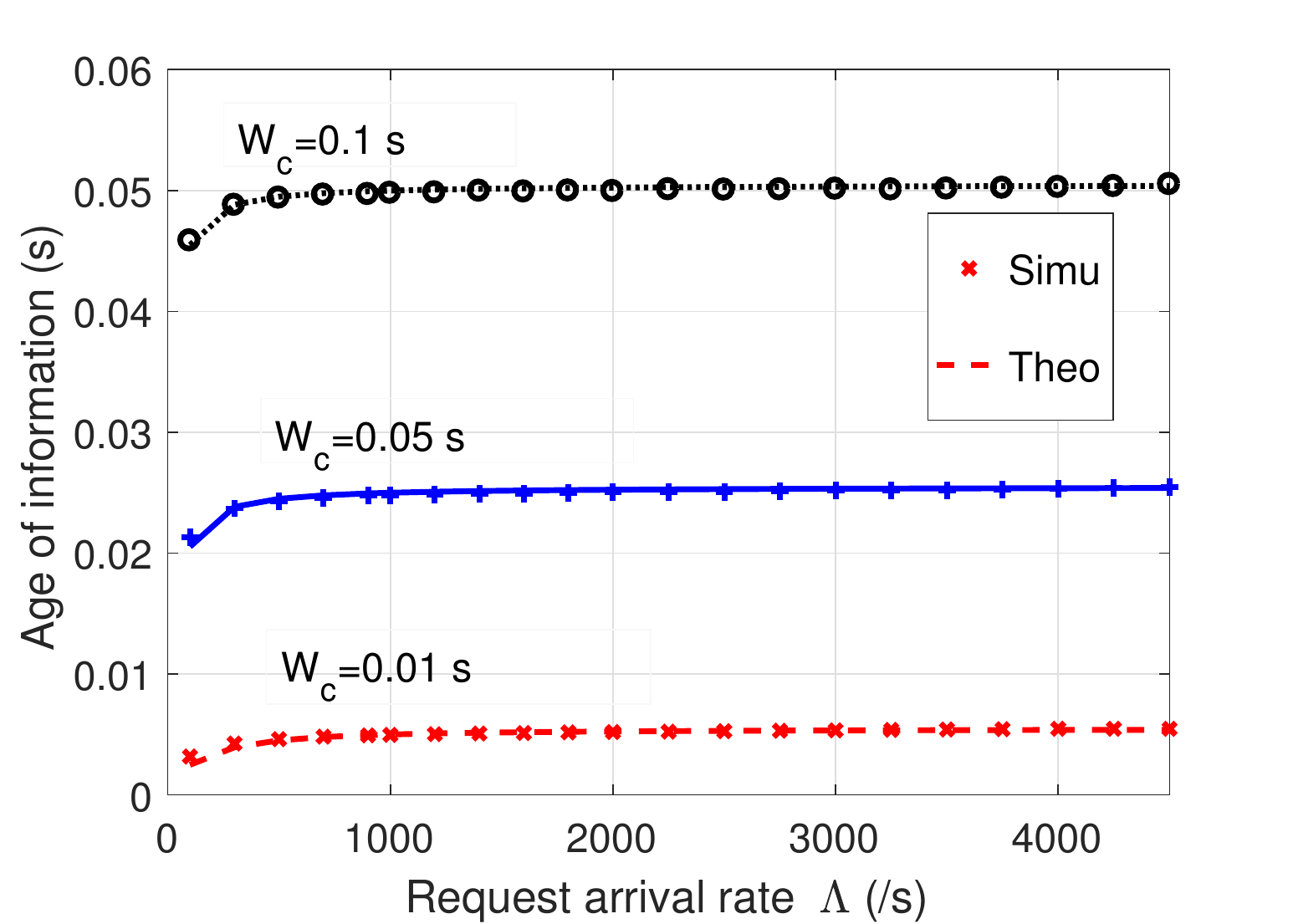}}\\
    		\subfloat[]{\includegraphics[width=3in]{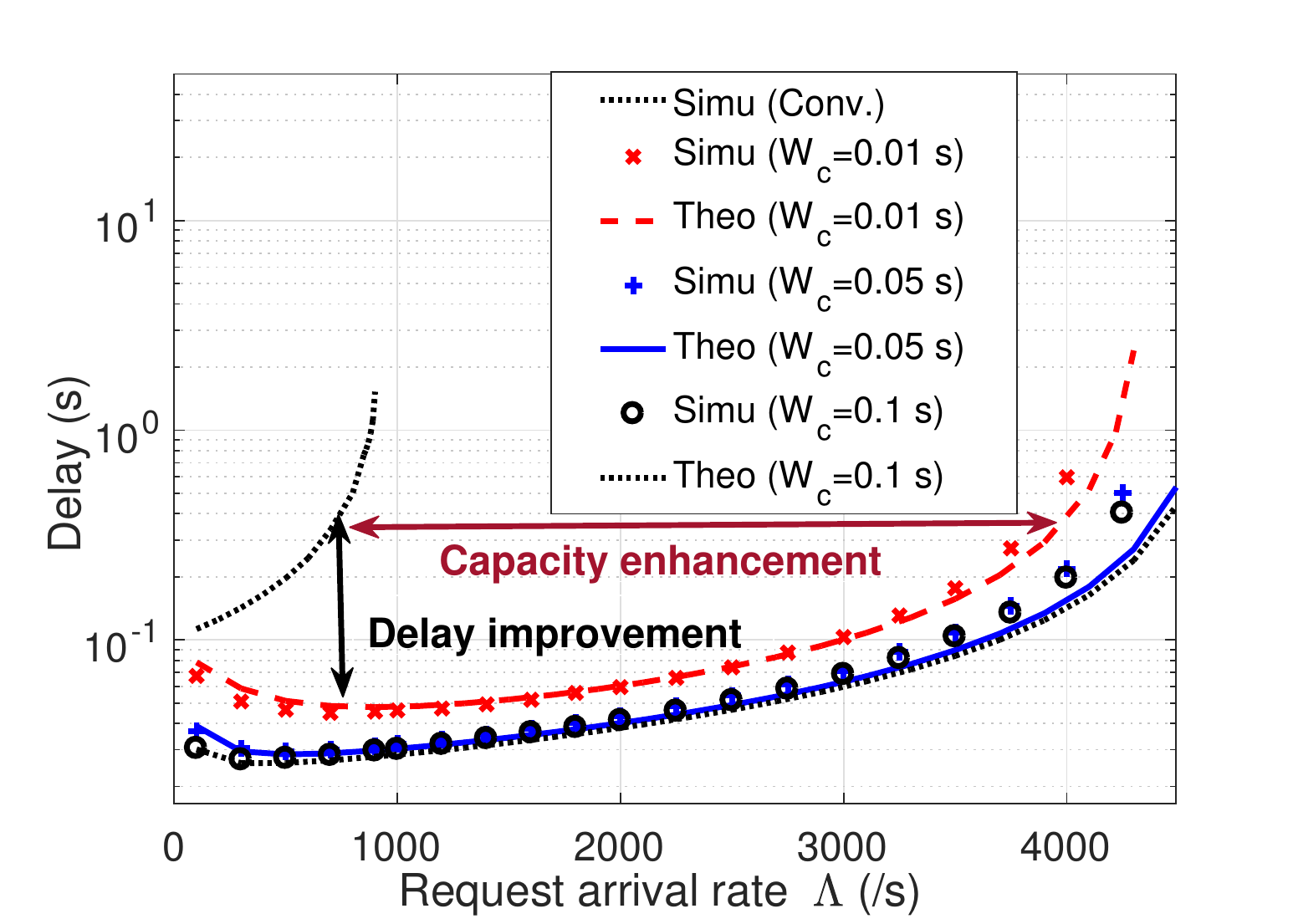}}\\
    		\subfloat[]{\includegraphics[width=3in]{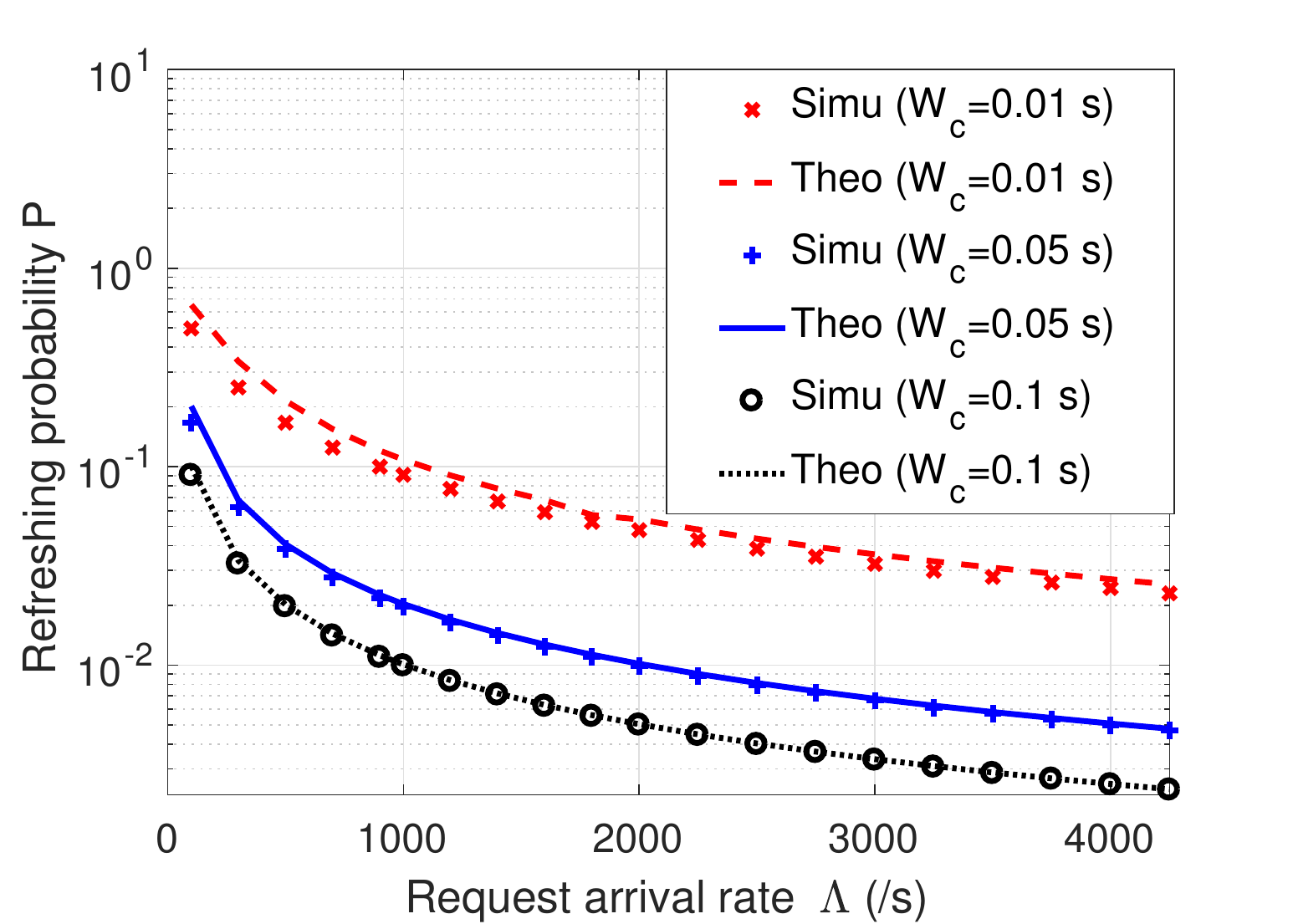}}
    		\caption{Analytical results validation with respect to request arrival rate in the single-source scenario, (a) Age of information, (b) Delay, (c) Refreshing probability.}
    		\label{fig_single_lambda}
    	\end{figure}

    	To begin with, simulations are conducted in the single-source scenario, where users are uniformly distributed within coverage and randomly raise requests.
    	In each simulation, the BS service process is simulated under the predefined refreshing window and request arrival rate.
    	The average AoI, delay, and refreshing probability are calculated based on 1 million request samples.
    	In addition, the theoretical results of average AoI, delay, and refreshing probability are obtained based on Theorems 1-3, which are compared with the simulation results in Figs.~\ref{fig_single_wc} and \ref{fig_single_lambda}.
    	The simulation and theoretical results are shown to be quit close in general, validating the approximated analysis. 
    	Furthermore, the results reveal the influence of key parameters.
    	
    	The influence of refreshing window is demonstrated in Fig.~\ref{fig_single_wc}.
    	Figure~\ref{fig_single_wc}(a) shows the average AoI increases with the refreshing window size in an approximated linear manner as the refresh window increases.
    	This result is consistent with the analytical results based on Eqs.~(\ref{eq_A_Win_first}) and (\ref{eq_A_Win_second}).
    	Figure~\ref{fig_single_wc}(b) shows that the average delay first decreases and then levels off with the refreshing window size, which is consistent with the analysis of Eqs.~(\ref{eq_D_max}) and (\ref{eq_D_min}). 
    	The reason can be explained by Fig.~\ref{fig_single_wc}(c).
    	As the refreshing window increases, the probability that a request triggers cache refreshing decreases, reducing the total transmission load and the average delay. 
    	When the refreshing window is sufficiently high, the cache is merely refreshed. 
    	In this case, the system transmission load achieves its minimum, and increasing refreshing window cannot further improve the delay performance.
    	
    	Figures~\ref{fig_single_wc}(a) and (b) reveal a tradeoff relationship between AoI and delay with respect to the refreshing window size.
    	Therefore, the proposed scheme can balance the AoI and delay performance on demand of the applications, by setting appropriate refreshing window size.
    	Furthermore, the AoI and delay performance are quite close under different request arrival rates, whereas the refreshing probability varies significantly. 
    	The important insight is that the proposed scheme can restrain the BS from frequently refreshing the same content item of high request rate. 
    	This helps to relieve traffic congestion and thus enhances the service capability of networks.
    	More details are provided in Fig.~\ref{fig_single_lambda}.
    	
    	The influence of request arrival rate is also evaluated, as shown in Fig.~\ref{fig_single_lambda}.
    	According to Fig.~\ref{fig_single_lambda}(a), the AoI increases with the request arrival rate concavely, and finally levels off at around half of the refreshing window size.
    	This result is consistent with Eqs.~(\ref{eq_A_lambda_1},\ref{eq_A_lambda_2},\ref{eq_A_lambda_3},\ref{eq_A_lambda_4}).
    	Notice that the average AoI is mainly influenced by the refreshing window instead of the request arrival rate.
    	Therefore, the proposed scheme can guarantee the content freshness under different load conditions, by setting appropriate refreshing window size.
    	The reason is explained in Fig.~\ref{fig_single_lambda}(c), where the refreshing probability is shown to decrease with the request arrival rate. Specifically, each request triggers cache refreshing with high probability, when the request arrival rate is extremely low. 
    	In this case, the AoI mainly depends on the interval between two successive user requests rather than the refreshing window size. 
    	Unlikely, the cache is refreshed almost periodically with the cycle length equals to the refreshing window size, under the heavy traffic load. 
    	In this case, the AoI mainly depends on the refreshing window size and does not change with the request arrival rate.
    	Therefore, the proposed scheme restrains the frequent cache refreshing at high request arrival rate, which is consistent with the results of Fig.~\ref{fig_single_wc}.

    	The average delay is shown as a convex function of request arrival rate, which firstly decreases and then increases.
    	This trend can be explained by Eq.~(\ref{eq_delay_mean_single}) of Theorem~3.
    	According to Eq.~(\ref{eq_delay_mean_single}), the average delay increases with both the refreshing probability $p$ and request arrival rate $\Lambda$.
    	However, the refreshing probability decreases with the request arrival rate as shown in Fig.~\ref{fig_single_lambda}(c), and thus the average delay does not always increase with the request arrival rate.
    	The results of Fig.~\ref{fig_single_lambda}(b) indicate that the variation of delay is dominated by the refreshing probability at low request arrival rate.
    	In addition, the average delay of conventional scheme is also illustrated as the dotted dash line, where the cache is refreshed with probability 1 (i.e., the eager refreshing scheme proposed for wired networks \cite{Si97_adaptive_refresh}).
    	The results show that the proposed scheme can effectively reduce the average delay and increase the BS capacity compared with the conventional scheme.
    	For example, if the average delay requirement is 1 second, the capacity (the maximal request arrival rate) is less than 1000 /s under the conventional scheme.
    	In comparison, the proposed scheme can increase the capacity to 4000 /s with $W_c$ set to 0.01s, wherein the average AoI of user received content is around 5 ms.
    	Furthermore, Fig.~\ref{fig_single_lambda}(b) shows that the capacity can be enhanced by enlarging the refreshing window but presents a marginal gain, which is consistent with Fig.~\ref{fig_single_wc}(b).
    	
    	\begin{figure}[!t]
    		\centering
    		\subfloat[] {\includegraphics[width=3in]{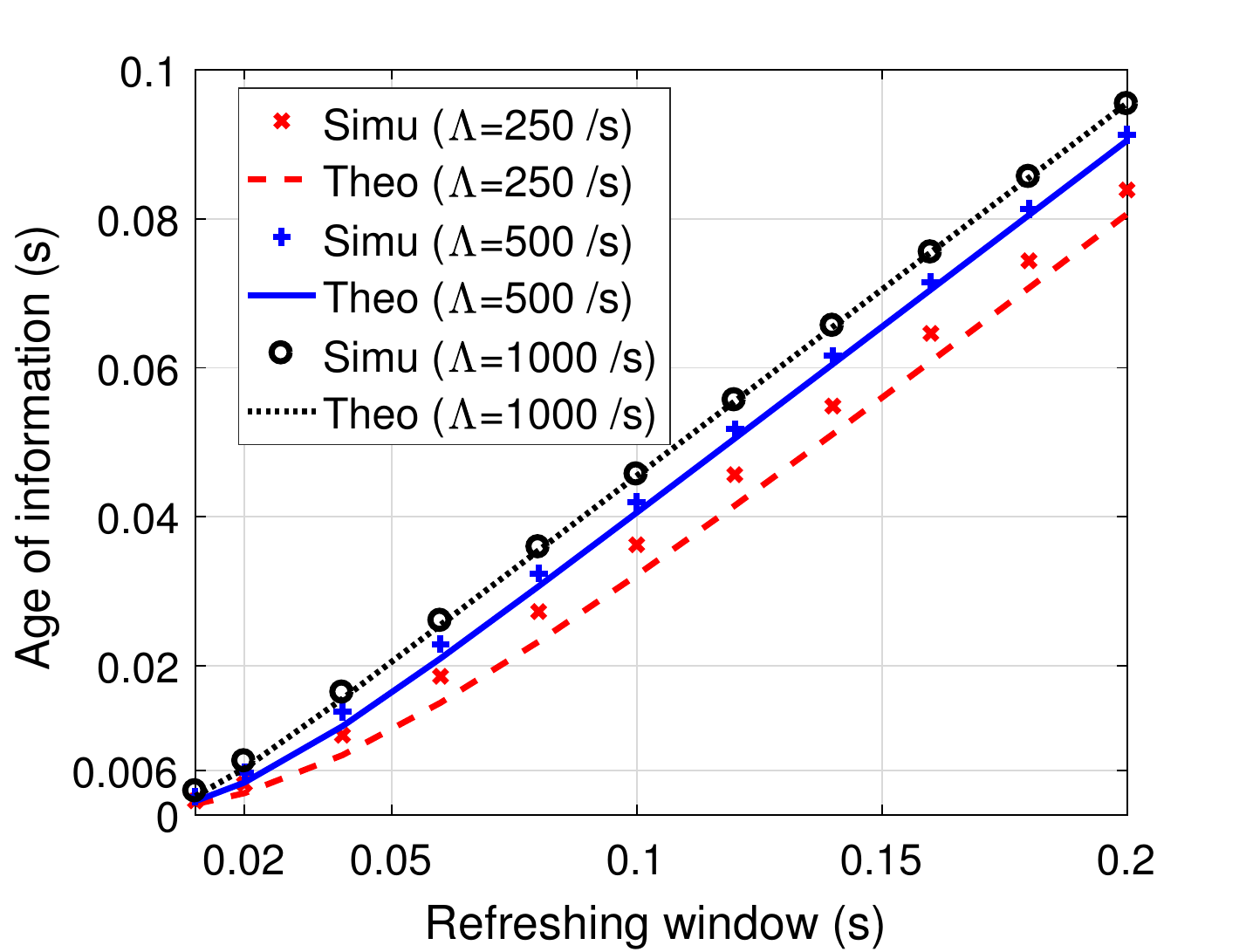}}
    		\subfloat[]{\includegraphics[width=3in]{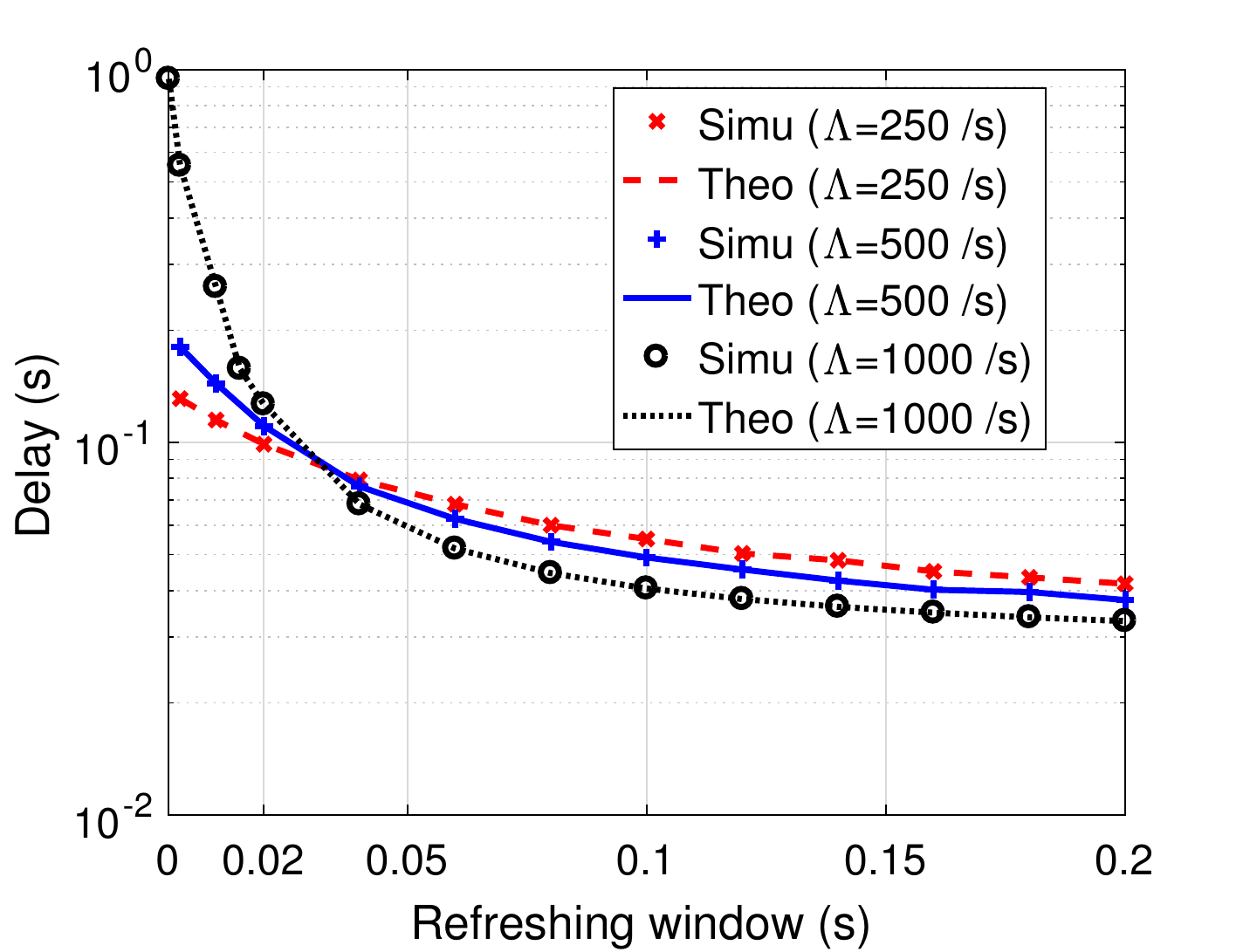}}
    		\caption{Analytical results validation in the multi-source scenario with uniform popularity, (a) Age of information, (b) Delay.}
    		\label{fig_multi_wc}
    	\end{figure}
    	
    	\subsection{Validation of Multi-Source Analysis}	
    	
    	Simulations are also conducted in multi-source scenario, where the content popularity is considered to follow Zipf distribution.
    	Without losing generality, the request probability of the $c$-th item is set to
    	\begin{align}
    	q_c=\frac{1/c^\nu}{\sum_{s=1}^C(1/s^\nu)},
    	\end{align}
    	where $\nu$ is a parameter reflecting the request concentration. 
    	Content items share the same request probability $1/C$ when $\nu=0$, i.e., uniform popularity.
    	The typical value for video-type service is $\nu=0.56$ \cite{Gill07_youtube}.
    	Simulations are conducted in case of ten items with uniform popularity under different refreshing window and request arrival rates, and the results are shown in Fig.~\ref{fig_multi_wc}.
    	The simulation and theoretical results are shown to be quite close, validating the approximated analysis.
    	Similar to the results of single-source scenario, the average AoI increases with the refreshing window in an approximately linear manner.
    	Accordingly, the content freshness can be well guaranteed by setting appropriate refreshing window regardless of the traffic load and content popularity.
    	In addition, the average delay decreases with the refreshing window convexly, revealing the AoI-delay tradeoff in the multi-case scenario.
    	The reason is that cache refreshing introduces more transmissions with more source nodes, which will be discussed in details later.
    	
    	\begin{figure}[t]
    		\centering
    		\subfloat[]{\includegraphics[width=1.5in]{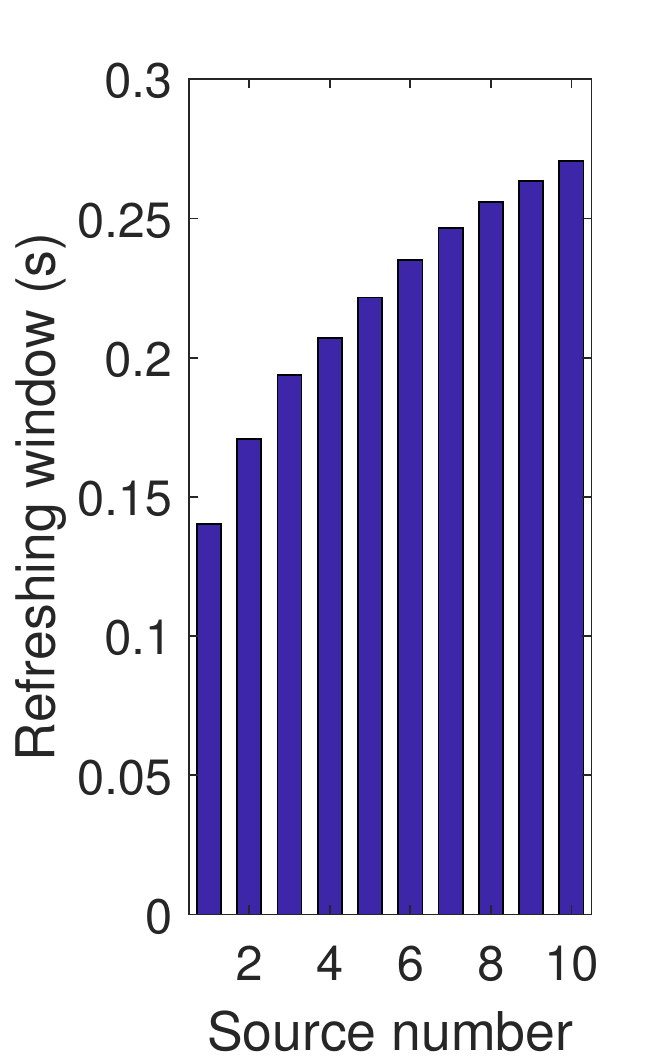}}
    		\hspace*{0.2in}
    		\subfloat[]{\includegraphics[width=1.5in]{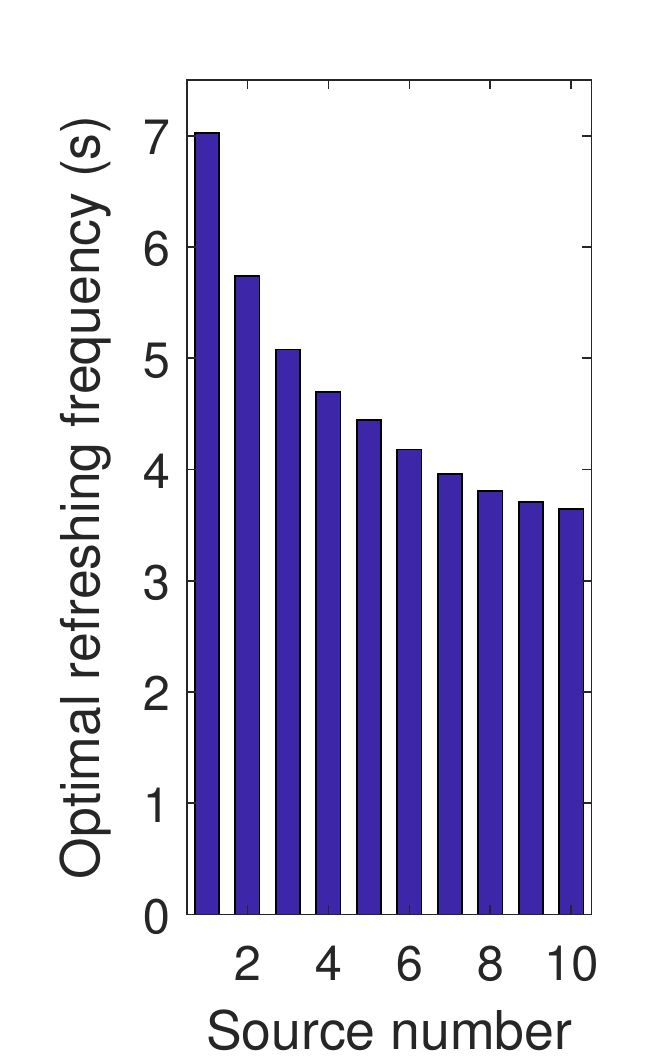}}
    		\hspace*{0.2in}
    		\subfloat[]{\includegraphics[width=1.5in]{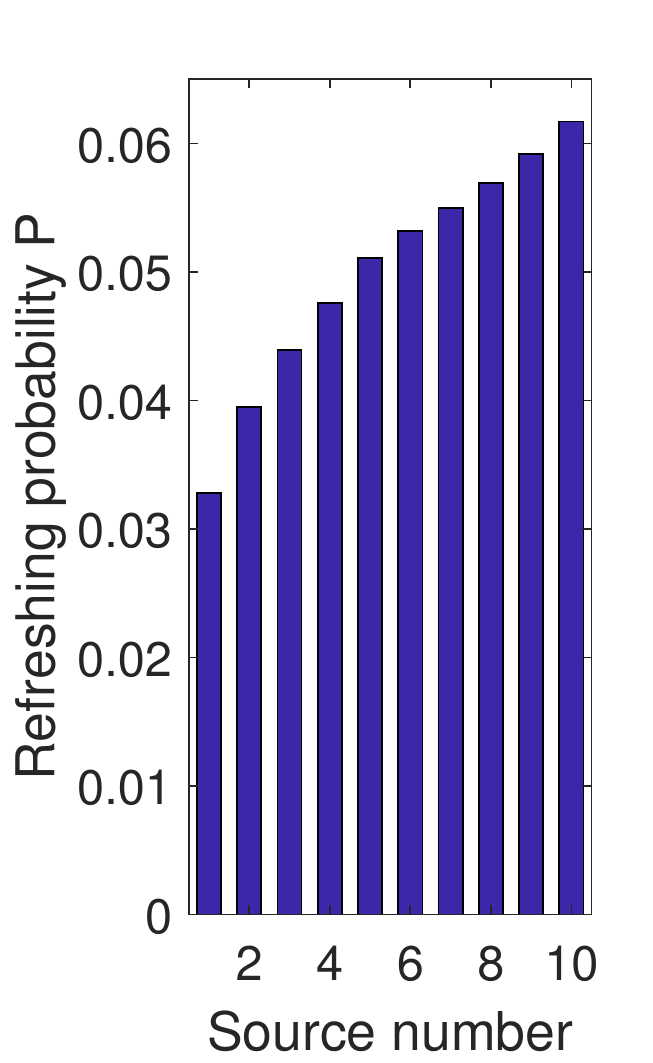}}
    		\caption{{Optimal refreshing of individual source: (a) Refreshing window, (b) Optimal refreshing frequency, (c) Refreshing probability, average AoI requirement 100 ms, sum request arrival rate $\Lambda$ = 2000 /s, Zipf exponent $\nu$ = 0.56.}}
    		\label{fig_multi_opt}
    	\end{figure}

    	\begin{figure}[t]
    		\centering
    		\includegraphics[width=3in]{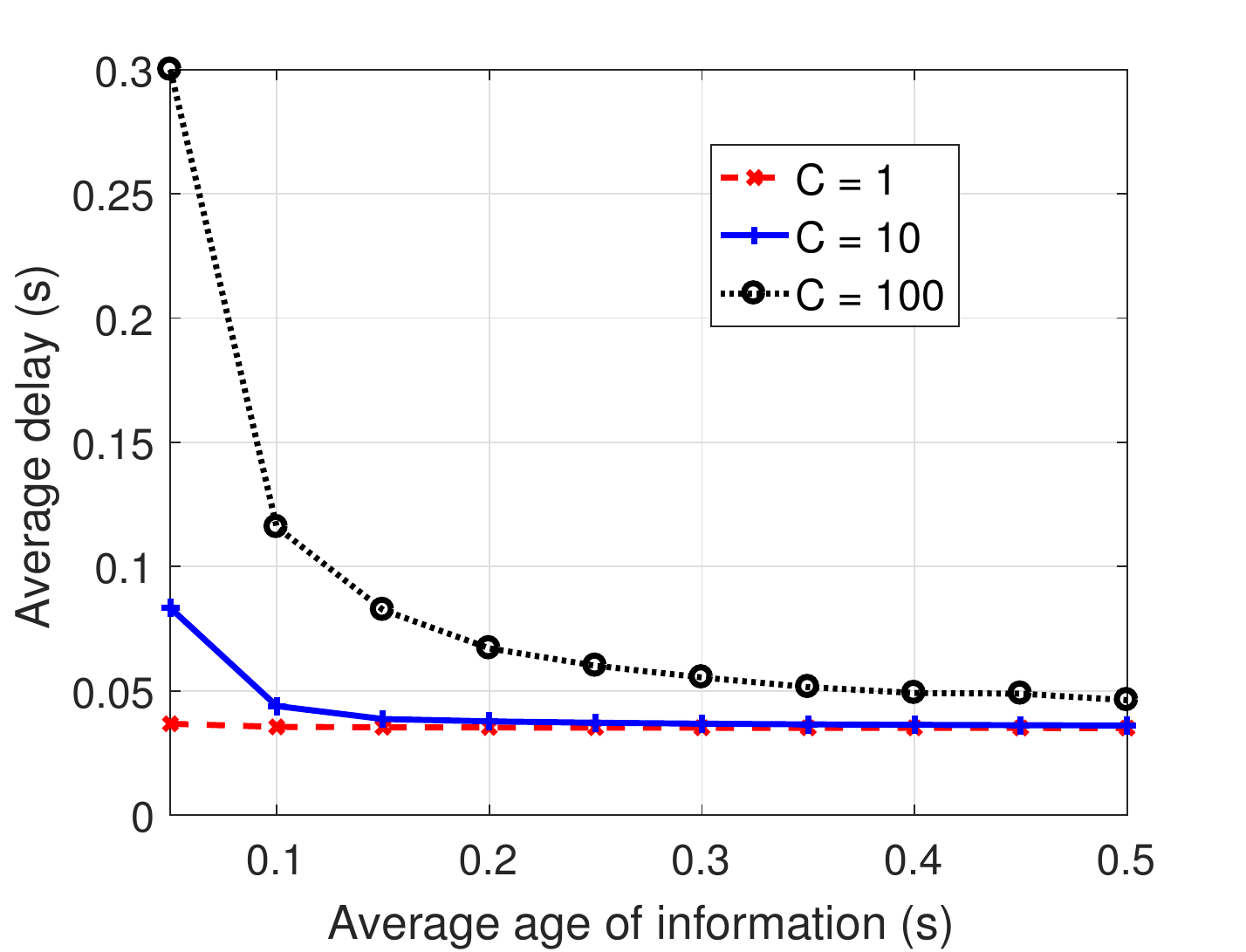}
    		\caption{Influence of item number on AoI and delay performance, sum request arrival rate $\Lambda$ = 2000 /s,
    			Zipf exponent $\nu$ = 0.56.}
    		\label{fig_multi_number}
    	\end{figure}
    	
    	\begin{figure}[t]
    		\centering
    		\includegraphics[width=3in]{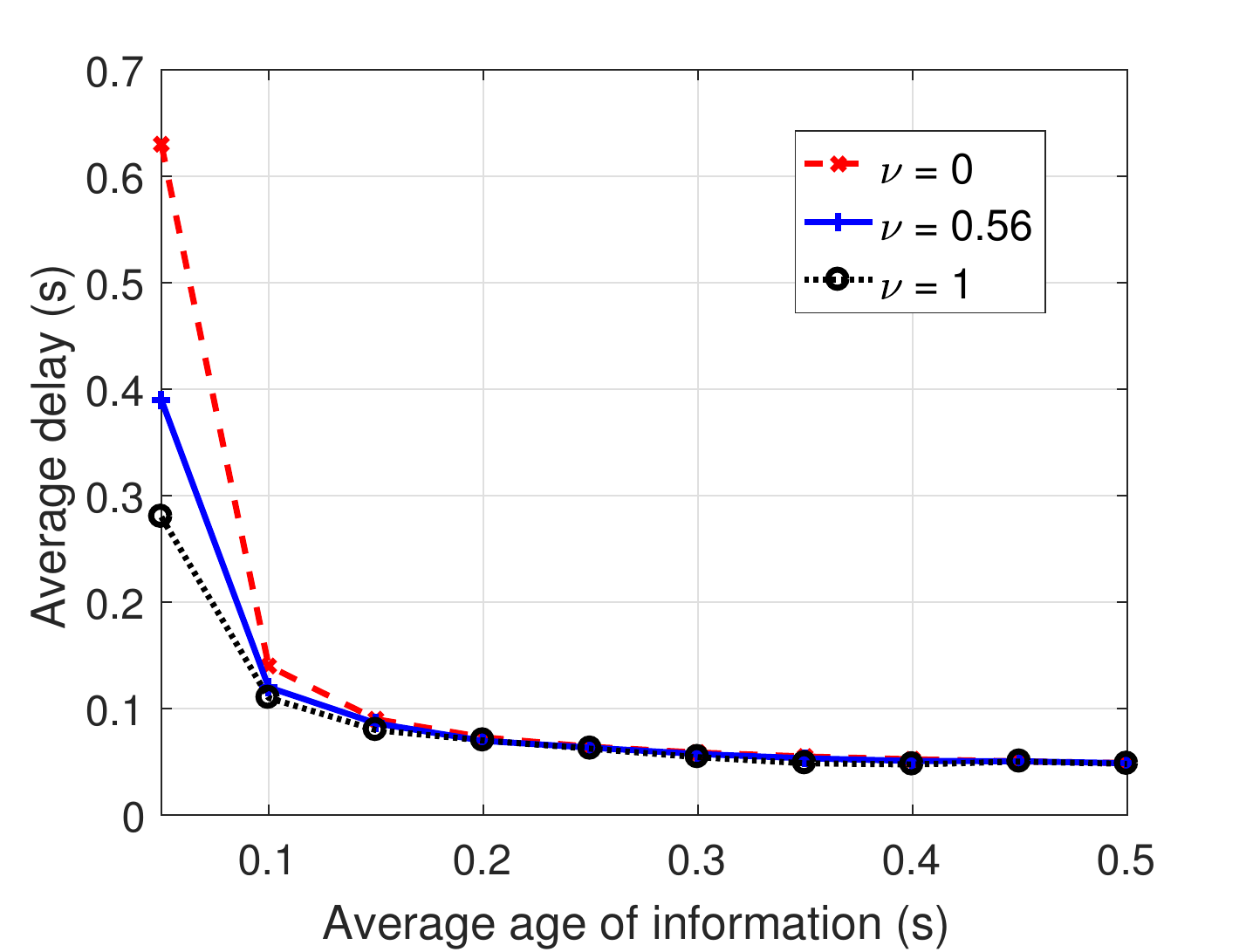}
    		\caption{Influence of content popularity distribution on AoI and delay performance, item number $C$ = 10, sum request arrival rate $\Lambda$ = 2000 /s.}
    		\label{fig_multi_popularity}
    	\end{figure}
    	
    	\subsection{Refreshing Window Optimization}
    	
    	In case of nonuniform popularity, the refreshing window sizes of each content should be optimized to enhance the system-level performance.
    	As an illustration, we consider 10 items whose popularity follows Zipf distribution of parameter 0.56, and the average AoI is required to be no larger than 100 ms.
    	The refreshing window can be optimized by solving (P1) with MATLAB convex optimization toolbox, and the results are shown in Fig.~\ref{fig_multi_opt}.
    	According to Fig.~\ref{fig_multi_opt}(a) and (b), the popular content items should be set with a smaller refreshing window and refreshed more frequently.
    	However, the refreshing probability of the popular content items are rather lower according to Fig.~\ref{fig_multi_opt}(c), indicating that the proposed scheme still retains the BS from frequently refreshing popular items.
    	In this way, the average AoI and delay is balanced.
    	
    	The influences of item number and request concentration are also investigated, as illustrated in Fig.~\ref{fig_multi_number} and Fig.~\ref{fig_multi_popularity}, respectively.
    	The delay performance is shown to degrade as the number of source nodes $C$ increases, according to Fig.~\ref{fig_multi_number}.
    	The reason is that the refreshing probability of each content item is increased, introducing more transmissions.
    	The insight is that the cost of maintaining content freshness increases with the cache size.
    	Therefore, the cache size and content placement should be optimized considering the requirement of content freshness in practice.
    	Furthermore, the average delay is shown to decrease with the concentration parameter $\nu$ in Fig.~\ref{fig_multi_popularity}.
    	This result indicates that mobile edge caching can provision better performance if the content requests are more concentrated, which is consistent with the case of static content items.
    	
    	\begin{figure}[t!]
    		\centering
    		\subfloat[]{\includegraphics[width=3in]{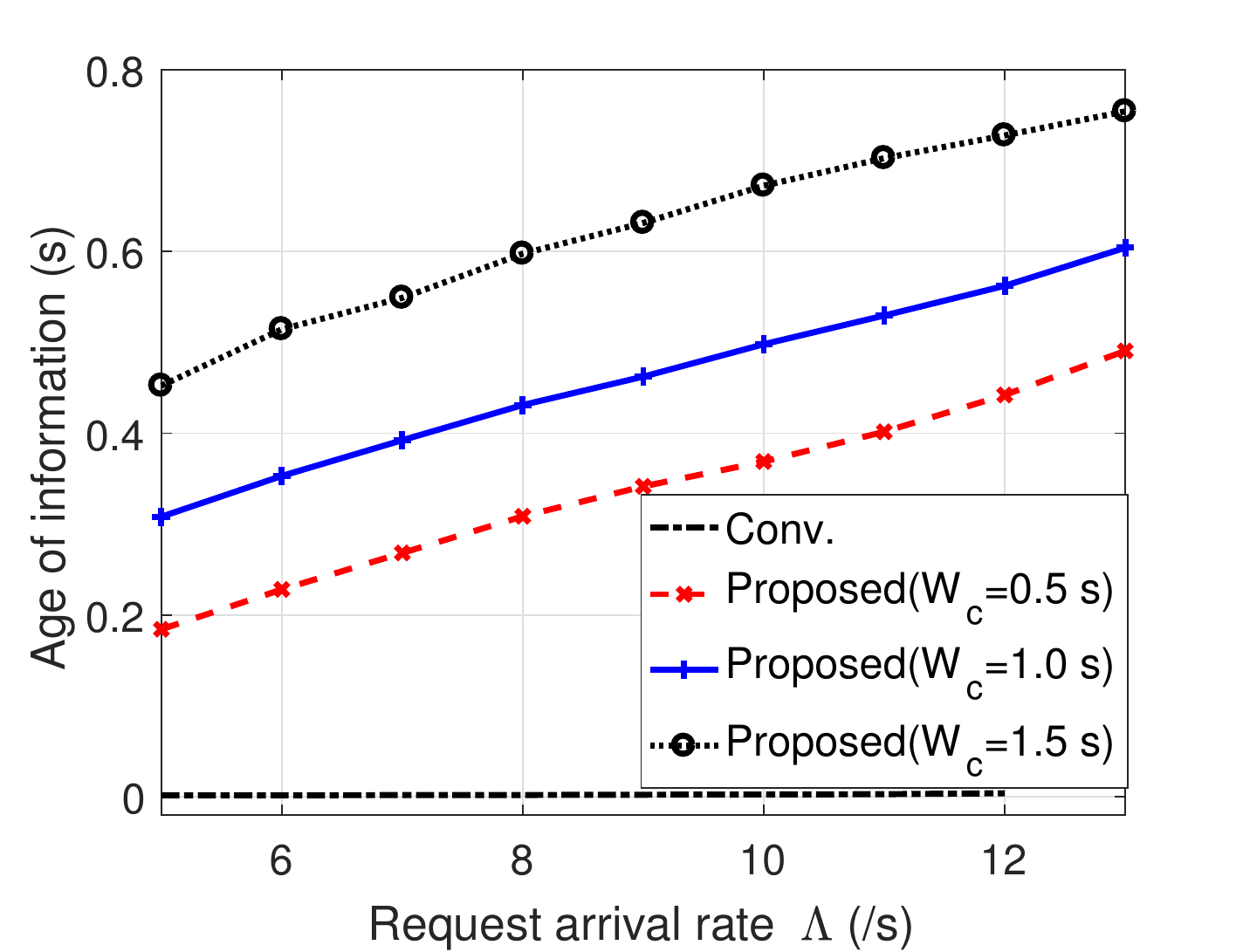}}
    		\subfloat[]{\includegraphics[width=3in]{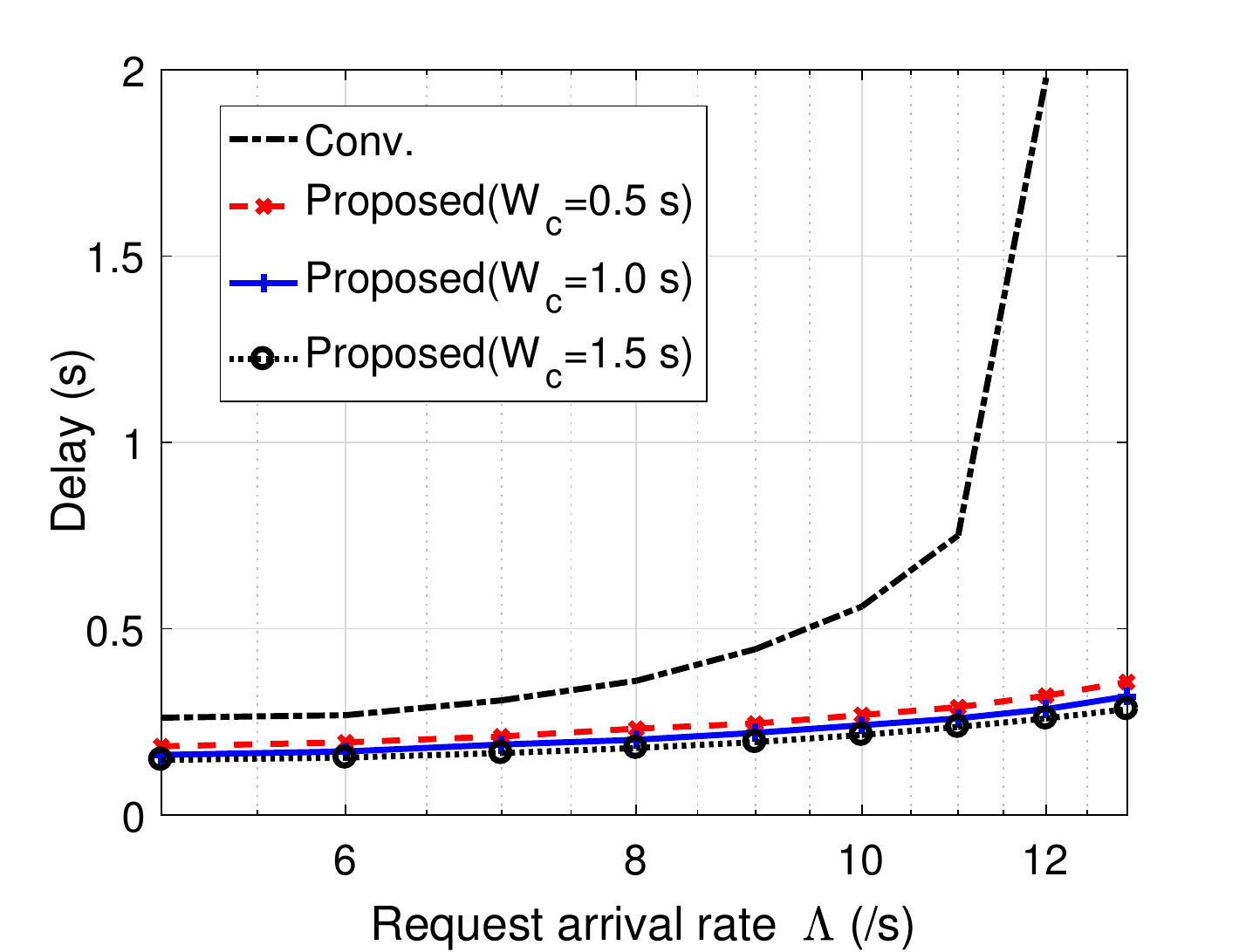}}
    		\caption{Performance evaluation with mobility based on Veins, (a) AoI, (b) Delay.}
    		\label{fig_veins}
    	\end{figure}
    	
    	\subsection{Real-Trace Mobility Simulations}
    	
    	A simulation platform is also built based on Veins to study the influence of user mobility in practical urban scenarios, using the map of Erlangen, Germany \cite{Veins}.
    	A cache-enabled BS is deployed at a road intersection, providing content service to vehicle users within 200 m. 
    	The locations of users are generated based on the Veins framework to reflect the real-trace vehicle mobility.
    	100 source nodes are randomly distributed within coverage, and the popularity follows Zipf distribution with $\nu=0.56$.	
    	The BS conducts the proposed freshness-aware content refreshing scheme to balance delay and AoI performances.
    	As comparison, the conventional scheme is adopted as a baseline, which always refreshing the cache to minimize the AoI.	
    	The simulation results of average AoI and delay are show in Fig.~\ref{fig_veins}.
    	The results show that the proposed scheme can effectively reduce the average delay compared with the conventional scheme, especially at high request arrival rates, by sacrificing the content freshness.
    	For example, the average delay is 2 second under the conventional scheme when the request arrival rate is 12 /s.
    	The average delay can be reduced by 85\% under the proposed scheme, by setting the refreshing window to 0.5 s.
    	Accordingly, the AoI will increase to around 420 ms, which is acceptable for many applications in practice, eg., the traffic congestion, real-time road map, the availability of gas stations and parking lots.

\section{Conclusions and Future Work}
    \label{sec_conclusions}

	This work has proposed a freshness-aware content refreshing scheme for mobile edge caching systems, where the BS refreshes the cached content items based on AoI upon user requests.
	The average AoI and service delay have been derived in closed forms approximately, demonstrating a tradeoff relationship with respect to the refreshing window size.
	Specifically, the average AoI of user-received content increases with the refreshing window in an asymptotically linear relationship if the system is not overloaded, while the average delay decreases in a convex manner.
	In the case of nonuniform content popularity, the refreshing window has been optimized for individual items to minimize the average delay while guaranteeing content freshness.
	Numerical results suggest to set a smaller refreshing window for popular content items, providing a guideline of the freshness-delay-optimized cache management in practice.
	For the future works, BSs can adopt the broadcast mode to further enhance the content delivery efficiency.
	



\bibliographystyle{IEEEtran}


\end{document}